\documentclass[conference,compsoc]{IEEEtran}

\usepackage{amsthm}  
\newtheorem{Definition}{Definition}
\ifdefined\theorem
\else
  \newtheorem{theorem}{Theorem}
\fi

\usepackage[T1]{fontenc}

\usepackage[utf8]{inputenc}
\usepackage{makecell}  
\usepackage{array}
\usepackage{tabularx}

\usepackage{threeparttable}
\usepackage{soul}	 
\AtBeginDocument{%
  }
    
\usepackage{tikz} 
\usepackage{setspace} 
\newcommand*{\circled}[1]{\lower.7ex\hbox{\tikz\draw (0pt, 0pt)%
    circle (.4em) node {\makebox[0.4em][c]{\small #1}};}}
\usepackage{mathtools}

\usepackage{mathrsfs}
\usepackage{filecontents}
\usepackage{color}
\usepackage{booktabs}   
\usepackage{graphicx}        
\usepackage{comment}
\usepackage[colorlinks,linkcolor=red,anchorcolor=blue,citecolor=blue]{hyperref}
\usepackage{textcomp}
\usepackage{url}
\usepackage{cite}
\newcommand{\etal}{\textit{et al}.} 
\usepackage{pifont}
\usepackage{marginnote}

\newcommand{\cmark}{\textcolor{black}{\ding{51}}}%
\newcommand{\xmark}{\textcolor{black}{\ding{55}}}%
\newcommand*\emptycirc[1][1ex]{\tikz\draw[thick] (0,0) circle (3pt);}
\newcommand*\halfcircc[1][1ex]{\tikz{
    \draw[thick] (0,0) circle (4pt);
    \fill[black] (0,0) -- ++(0, 4pt) arc[start angle=90, end angle=-90, radius=4pt] -- cycle;
}} 
\newcommand*\halfcirc[1][1ex]{\tikz{
    \draw[thick] (0,0) circle (5pt);
    \fill[black] (0,0) -- ++(0, 5pt) arc[start angle=90, end angle=-90, radius=5pt] -- cycle;
}} 
\usepackage{tabularx}
\usepackage{stmaryrd,amsmath,amssymb}
\usepackage{adjustbox} 
\usepackage{multirow, booktabs, diagbox}
\usepackage{tabularx}
\usepackage{algorithm}
\usepackage{algorithmicx}
\usepackage{algpseudocode}
\usepackage{enumitem}
\usepackage[most]{tcolorbox} 
\newtcolorbox{mybox}[2][]{%
  enhanced,
  title        = {#2},
  attach boxed title to top left={xshift=+3mm,yshift*=-3mm},
  colback      = white,
  colframe     = black,
  fonttitle    = \bfseries,
  fontupper    = \small,
  fontlower    = \small,
  colbacktitle = black!3!white,
  coltitle     = black,
  #1
}
\newtcbox{\xmybox}[1][red]{on line, arc=7pt,colback=#1!10!white,colframe=#1!50!black, before upper={\rule[-3pt]{0pt}{10pt}},boxrule=1pt, boxsep=0pt,left=6pt,right=6pt,top=1pt,bottom=1pt}
\newcommand{\mycaptionspace}{\vspace{-1.8mm}}

\usepackage{tikz} 

\usetikzlibrary{fit,calc}

\colorlet{pink}{red!20}
\definecolor{blue0}{HTML}{C0D3E3}
\definecolor{blue1}{HTML}{9299C1}
\definecolor{blue2}{HTML}{9170AB}
\floatname{algorithm}{Algorithm}

\newcommand{\sysname}{\textsf{FLAME}} 
\newcommand{\red}[1]{\textcolor{red}{#1}}

\begin{document}
\title{$\sysname$: Flexible and Lightweight Biometric Authentication Scheme in Malicious Environments 
}

\author{\IEEEauthorblockN{Fuyi Wang}
\IEEEauthorblockA{\textit{Singapore University of Technology} \\
\textit{and Design} \\ 
wong$\_$fuyi@outlook.com}
\and
\IEEEauthorblockN{Fangyuan Sun}
\IEEEauthorblockA{\textit{Qingdao} \\ 
\textit{University} \\ 
sakgofish@gmail.com}
\and
\IEEEauthorblockN{Mingyuan Fan}
\IEEEauthorblockA{\textit{East China Normal} \\  
\textit{ University} \\ 
fmy2660966@gmail.com}
\and
\IEEEauthorblockN{Jianying Zhou}
\IEEEauthorblockA{\textit{Singapore University of Technology } \\
\textit{and Design} \\ 
jianying\_zhou@sutd.edu.sg} 
\and 
\IEEEauthorblockN{Jin Ma}
\IEEEauthorblockA{\textit{Clemson University
} \\ 
jin7@clemson.edu}
\and
\IEEEauthorblockN{Chao Chen}
\IEEEauthorblockA{\textit{RMIT University} \\ 
chao.chen@rmit.edu.au}
\and
\IEEEauthorblockN{Jiangang Shu}
\IEEEauthorblockA{\textit{Guangzhou University} \\ 
jiangangshu@gmail.com}
\and
\IEEEauthorblockN{Leo Yu Zhang }
\IEEEauthorblockA{\textit{Griffith University} \\ 
leo.zhang@griffith.edu.au}
} 

\maketitle             

\begin{abstract}
Privacy-preserving biometric authentication (PPBA) enables client authentication without revealing sensitive biometric data, addressing privacy and security concerns.
Many studies have proposed efficient cryptographic solutions to this problem based on secure multi-party computation, typically assuming a semi-honest adversary model, where all parties follow the protocol but may try to learn additional information.
However, this assumption often falls short in real-world scenarios, where adversaries may behave maliciously and actively deviate from the protocol. 

In this paper, we propose, implement, and evaluate $\sysname$, a \underline{F}lexible and \underline{L}ightweight biometric \underline{A}uthentication scheme designed for a \underline{M}alicious \underline{E}nvironment. 
By hybridizing lightweight secret-sharing-family primitives within two-party computation, $\sysname$ carefully designs a line of supporting protocols that incorporate integrity checks with rationally extra overhead. 
Additionally, $\sysname$ enables server-side authentication with various similarity metrics through a cross-metric-compatible design, enhancing flexibility and robustness without requiring any changes to the server-side process.
A rigorous theoretical analysis validates the correctness, security, and efficiency of $\sysname$.
Extensive experiments highlight $\sysname$'s superior efficiency, with a communication reduction by {$97.61\times \sim 110.13\times$} and a speedup of {$ 2.72\times \sim 2.82\times$ (resp. $ 6.58\times \sim 8.51\times$)} in a LAN (resp. WAN) environment, when compared to the state-of-the-art work. 
\end{abstract}

\section{Introduction}
\label{sec:intro}
Biometric authentication offers a convenient way to authenticate clients' identities based on their unique biological traits, eliminating the need to memorize passwords or manage other secret credentials \cite{cheng2024nomadic}. 
To decouple authentication from a specific device, biometric templates are often outsourced to cloud servers rather than stored locally.
This enables clients to authenticate seamlessly across devices, enhancing scalability and flexibility. 
However, this raises significant privacy and security concerns, as biometric templates are highly private and sensitive. Additionally, unlike passwords, biometric templates---such as fingerprints, facial images, iris scans, and genomic data---cannot be superseded if compromised by cloud servers or other adversaries.  
Therefore, privacy-preserving biometric authentication (PPBA) schemes \cite{nieminen2020practical,wei2020privacy,pradel2021privacy,huang2023efficient,al2024secure} have emerged to strengthen the protection of biometric templates. They leverage cryptographic primitives such as homomorphic encryption (HE) \cite{paillier1999public} and/or multi-party computation (MPC) \cite{goldreich2019play},  including garbled circuit (GC) and secret sharing (SS) techniques.
PPBA schemes are classified according to their underlying security models, namely the semi-honest and malicious settings. A comprehensive comparison of representative schemes under both models is presented in Table~\ref{tab:comparison}.

\begin{table*}[!t]
\caption{Comparison of PPBA schemes. }
\vspace{-3mm}
\centering
\begin{adjustbox}{width=0.93\textwidth,center}
\begin{tabular}{ccccccccc} \\ \hline \hline
Schemes & Model & Primitives & Security & Integrity & $f^{\text{lin}}$  & $f^{\text{nonlin}}$  & Metrics & Latency$^2$ \\ \hline
Nieminen~\etal~\cite{nieminen2020practical} &   Client-server  &  HE,GC  & \tikz \draw (0,0) circle (5pt); & \xmark  & \xmark & \xmark$^1$ & Euclidean   & High \\
Wei~\etal~\cite{wei2020privacy}   &   Client-server &  HE      & \tikz \draw (0,0) circle (5pt); & \xmark  & \xmark & \xmark$^1$ & Cosine       & High \\
Pradel~\etal~\cite{pradel2021privacy}   & Client-server & HE   & \tikz \draw (0,0) circle (5pt); & \xmark   & \xmark & \xmark$^1$ & Euclidean   & High \\
Huang~\etal~\cite{huang2023efficient} & Two servers  & HE,GC & \tikz \draw (0,0) circle (5pt); & \xmark  & \xmark & \xmark$^1$ & Euclidean   & High \\
Wang~\etal~\cite{wang2025SEBioID}   & Two servers &  SS,GC  & \tikz \draw (0,0) circle (5pt); & \xmark  & \cmark & \xmark\textcolor{white}{$^1$}  & Euclidean   & Moderate \\
Im~\etal~\cite{im2020practical}   & Client-server & HE & \halfcirc & HE  & \xmark & \xmark$^1$ & Euclidean & Moderate \\
Bassit~\etal~\cite{bassit2021biometric}   & Client-server & HE & \halfcirc & ZKP  & \xmark & \xmark$^1$ & Likelihood  & Moderate \\
Al-Mannai~\etal~\cite{al2024secure} &  Client-server &  HE  & \halfcirc & ZKP & \xmark & \xmark{$^1$} & Euclidean     & Low \\
Cheng~\etal~\cite{cheng2024nomadic} & Two servers  & OptSS, FuncSS & \tikz \fill (0,0) circle (5pt); & MAC & \xmark & \cmark & Cosine      & Moderate \\
$\sysname$  &  Two servers  &  OptSS, FuncSS & \tikz \fill (0,0) circle (5pt); & MAC & \cmark  & \cmark & Cosine, Euclidean & Low \\ \hline \hline
\end{tabular}
\end{adjustbox} 
\begin{tablenotes}[flushleft]
\item \tikz \draw (0,0) circle (4pt);: semi-honest, $\halfcircc$: malicious client-only, \tikz \fill (0,0) circle (4pt);: malicious. HE: homomorphic encryption, SS: secret sharing, GC: garbled circuit, OptSS: optimized SS, FuncSS: function SS, $f^{\text{lin}}$ and $f^{\text{nonlin}}$: the optimization of linear and non-linear functions,
Cosine: Cosine similarity, Euclidean: squared Euclidean distance, Likelihood: log likelihood ratio classifier.
\item $^1$ These schemes optimize computation using precomputation and packing techniques, rather than optimizing the protocols.
\item $^2$ The latency is evaluated based on the results reported in their papers.
\end{tablenotes}
\vspace{-2mm}
\label{tab:comparison}
\end{table*}

\textbf{Related semi-honest schemes.} Most existing PPBA studies target the semi-honest setting. Studies \cite{pradel2021privacy,saraswat2022phbio,wei2020privacy} employ (fully/partially) HE to achieve privacy-preserving facial recognition. However, the significant computational and communication overhead associated with HE renders it impractical for real-world scenarios. To improve efficiency, many hybrid solutions that combine HE with GC \cite{jarvinen2019pilot,nieminen2020practical,huang2023efficient}, SS \cite{wang2025SEBioID}, or cancelable biometrics \cite{shahreza2022hybrid}, have been proposed for PPBA.  
While these hybrid solutions do achieve performance gains, HE-involved schemes still require the client to retain a private decryption key, which nullifies the advantage of biometric authentication which promotes authentication without the need to remember passwords or manage secret keys.
To address this, recent efforts have been shifted towards utilizing MPC primitives, which offer more efficient alternative authentication schemes for various biometrics, such as fingerprint \cite{yang2022privacy,wang2025SEBioID}, voice \cite{treiber2019privacy}, and iris \cite{fualuamacs2021assessment}. However, these semi-honest-centric studies assume that all parties follow protocols without deviation, which is still challenging to meet in reality \cite{blanton2024privacy}.

\textbf{Related malicious schemes.} PPBA schemes against malicious adversaries have recently attracted increasing attention. 
Broadly, two models have emerged: the client-server interactive model and the fully outsourced two-server model. In the client-server model, existing works typically assume a malicious client, and employ mechanisms such as commitments \cite{im2020practical} and the zero-knowledge proof (ZKP) \cite{bassit2021biometric,al2024secure} to ensure the integrity of intermediate results from the client. However, these approaches often require frequent client participation and fail to ensure the correctness of server-side computations, limiting their scalability and practicality for resource-constrained clients. 
To address these challenges, the literature \cite{bassit2021fast,al2024secure} introduced provably PPBA schemes that leverage ZKPs to detect and mitigate malicious behavior from servers under a fully outsourced setting. 
\cite{barni2019semba,cheng2024nomadic} utilized the SPDZ MPC protocol to learn the biometric authentication result in a secure manner, employing the lightweight message authentication code (MAC) to ensure result integrity.

Maliciously secure protocols are significantly more complex, with computational overhead often exceeding that of semi-honest counterparts by an order of magnitude. 
Since an equal amount of linear computations are significantly faster than non-linear ones over the secret-sharing domain, existing malicious-against computation schemes focus on optimizing costly non-linear comparisons \cite{cheng2024nomadic} or improving efficiency using packing techniques \cite{im2020practical,bassit2021biometric}, while largely overlooking improvements in linear computation efficiency. 
However, in PPBA, the linear similarity computation remains a major performance bottleneck due to the large volume and high dimensionality of reference biometric templates, leaving substantial room for optimization (\textbf{Challenge \circled{1}}). 
Moreover, these schemes rely on over-the-threshold Cosine similarity or squared Euclidean distance for biometric authentication, which introduces two extra challenges. \textbf{Challenge \circled{2}:} Determining the appropriate threshold is challenging due to its variability across different applications and environments, making it difficult to establish a one-size-fits-all solution. A poorly chosen threshold can lead to either high false acceptance rates or high false rejection rates, both of which compromise the effectiveness of the authentication system. 
\textbf{Challenge \circled{3}:} Different types of biometric traits, each with distinct characteristics, require different similarity metrics. These schemes typically support only a single similarity metric, which limits their flexibility and robustness. 
Switching to a different metric often requires corresponding modifications to the authentication protocol. For example, Cosine similarity requires the computed value to exceed a threshold for successful authentication, whereas Euclidean distance requires it to fall below a threshold.

\textbf{Our contributions.} To address these challenges, we design, implement, and evaluate $\sysname$, a fast, flexible, and secure biometric authentication scheme built on a two-party computation (2PC) setup. 
In $\sysname$, even if both servers access intermediate results, they cannot reconstruct the original biometric templates, thereby preserving client privacy. It further ensures malicious security under a dishonest majority, guaranteeing the integrity of authentication outcomes (e.g., preventing forged ``success'' or ``failure'' results).
Also, $\sysname$ is seamlessly compatible with both Cosine similarity and Euclidean distance metrics by bridging the semantic gap, ensuring flexible applicability across different biometric authentication techniques.
To boost efficiency, $\sysname$ integrates lightweight cryptographic primitives with an offline-online paradigm. It significantly reduces online authentication latency while achieving malicious security with only $\approx 2\times$ overhead of semi-honest PPBA schemes.
This enables real-time, large-scale biometric authentication with strong security guarantees.
In summary, our contributions are threefold.

\begin{itemize}[leftmargin=*,topsep=0pt]
    \item We present $\sysname$, a flexible and crypto-friendly biometric authentication scheme that supports diverse similarity metrics in the malicious setting. Based on a 2PC setup, one of the servers is always assumed to be malicious, ensuring strong security against internal adversaries.

    \item Based on secret-sharing-family primitives, we customize a suite of secure linear and non-linear protocols for PPBA by leveraging an offline-online paradigm. These protocols are carefully crafted to minimize online computational overhead and reduce communication to a single round. 
    
    \item We formally prove the correctness, efficiency, and security of $\sysname$. We conduct extensive experiments and compare our design with state-of-the-art studies. The results highlight that $\sysname$ achieves high efficiency and scalability, with face recognition serving as a representative application. 
\end{itemize}

\section{Secret Sharing}
\label{sec:Preliminaries}
\textbf{Optimized secret sharing} (OptSS) \cite{ben2019turbospeedz} is an enhanced arithmetic secret-sharing approach, dividing operations in the secret-sharing domain into pre-processing (i.e., offline) and online phases. In the pre-processing phase, correlated random offset shares ($\lambda$-values) are generated for the input and output wires of each gate of the arithmetic circuit. These input-independent shares can be executed prior to the online phase, where the actual function $f$ is evaluated via efficient secure computation.
This paper adopts the 2-out-of-2 OptSS over the ring $\mathbb{Z}_{2^{l}}$ and we now give the formal definition.
On the secret $\langle x\rangle \in \mathbb{Z}_{2^{l}}$ ($\langle x\rangle_0+ \langle x\rangle_1=x$), parties call $\prod_\textnormal{OptSS}(\langle x \rangle)$ to learn OptSS-based shares $[\![x]\!]_{\theta}=(\Delta_{x}, \langle \lambda_{x} \rangle_{\theta}) $ $\in \mathbb{Z}_{2^{l}}$, s.j. $\Delta_{x} =x+\lambda_{x}\pmod{2^l}$, where $\langle \lambda_{x} \rangle$ is the $x$-associated random offset and $\lambda_{x} =\sum_{\theta=0}^{1}\langle \lambda_{x} \rangle_{\theta}$. With $2$ shares $(\Delta_{x}, \langle \lambda_{x} \rangle_{0})$, $(\Delta_{x}, \langle \lambda_{x} \rangle_{1})$, the secret $x$ can be opened over the plaintext space $\mathbb{Z}$, i.e., $x=\Delta_{x}-\langle \lambda_{x} \rangle_0 - \langle \lambda_{x} \rangle_1 \pmod{2^l}$. 
Given one public value $p$ and two secrets $\langle x\rangle, \langle y\rangle$, where $\mathcal{P}_{\theta}$ $(\forall \theta \in \{0,1\})$ holds OptSS shares $(\Delta_{x}, \langle \lambda_{x} \rangle_{\theta})$ and $(\Delta_{y}, \langle \lambda_{y} \rangle_{\theta})$, four basic secure computations for addition (add) and multiplication (mult) are defined as follows. 
\begin{itemize}[leftmargin=*,topsep=0pt]

    \item[-] $\prod_{\textnormal{add}}\leftarrow[\![x]\!] + [\![y]\!]$: $\mathcal{P}_\theta$ computes $\Delta_{z}=\Delta_{x}+\Delta_{y}$ and $\langle \lambda_{z} \rangle_{\theta}=\langle \lambda_{x} \rangle_{\theta} +\langle \lambda_{y} \rangle_{\theta}$.
    \item[-] $[\![x]\!] + p$: $\mathcal{P}_\theta$ computes $\Delta_{z}=\Delta_{x}+p$ and $\langle \lambda_{z} \rangle_{\theta}=\langle \lambda_{x} \rangle_{\theta} $. 
    \item[-] $\prod_{\textnormal{mult}}\leftarrow[\![x]\!] \cdot [\![y]\!]$: {A shared multiplication triple $(\langle a \rangle, \langle b \rangle,\langle c \rangle)$, values $\delta_x$, $\delta_y$, $\langle \lambda_{z} \rangle$ are generated, where $c=ab$, $\delta_x=a-\lambda_x$,$\delta_y=b-\lambda_y$}. Two servers first compute the shared $\langle \Delta_{z} \rangle_{\theta}= \theta \cdot\left(\Delta_x+\delta_x\right)\left(\Delta_y+\delta_y\right)-\langle a\rangle_{\theta} \left(\Delta_y+\delta_y\right)-\left(\Delta_x+\delta_x\right)\langle b \rangle_{\theta}+\langle c \rangle_{\theta} +\langle \lambda_{z}\rangle_{\theta}$. $\forall \theta \in\{0,1\}$, $\mathcal{P}_{\theta}$ sends $\langle \Delta_{z} \rangle_{\theta}$ to $\mathcal{P}_{1-\theta}$ to open $\Delta_{z}$. Finally, the OptSS shares $(\Delta_{z}, \langle \lambda_{z} \rangle_{\theta})$ for the product of $\langle x \rangle$ and $\langle y \rangle$ is obtained.
     \item[-] $[\![x]\!] \cdot p$:  $\mathcal{P}_\theta$ computes $\Delta_{z}=\Delta_{x} \cdot p$ and $\langle \lambda_{z} \rangle_{\theta}=\langle \lambda_{x} \rangle_{\theta} $. 

\end{itemize}

\textbf{Authenticated secret sharing} (AuthSS) 
from $\textnormal{SPD}\mathbb{Z}_{2^l}$~\cite{cramer2018spd} is a cryptographic technique with the information-theoretic message authentication code (MAC) that ensures both the confidentiality and integrity of secret-shared values. It is a foundational building block in secure multiparty computation (MPC) protocols against malicious adversaries. 
Due to the presence of zero divisors in $\mathbb{Z}_{2^l}$, the parties must perform computations over a larger ring modulo $2^{l+s}$, while the security and correctness is only guaranteed modulo $2^l$. 
Specifically, assume each party holds a shared $\langle \phi \rangle \stackrel{\$}{\leftarrow} \mathbb{Z}_{2^{s}}$ of a secret global MAC key $\phi = \sum_{\theta=0}^{1}\langle \phi \rangle_\theta \pmod 2^{l+s}$.
Given a shared value $\langle x \rangle \in \mathbb{Z}_{2^{l+s}}$, each party $\mathcal{P}_\theta$ holds the authenticated share $\{\langle x \rangle, \langle m \rangle \in \mathbb{Z}_{2^{l+s}}, \langle \phi \rangle \in \mathbb{Z}_{2^{s}}\}$.  
These components satisfy $\sum_{\theta=0}^{1} \langle m \rangle_\theta \equiv_{l+s}\left(\sum_{\theta=0}^{1} \langle x \rangle_\theta\right) \cdot\left(\sum_{\theta=0}^{1} \langle \phi \rangle_\theta\right)$, where $\equiv_{l+s}$ denotes ring module $2^{l+s}$.
Similarly, for a function $f$, the output includes the shares $\langle f(x)\rangle $ and $\langle \phi f(x)\rangle \in \mathbb{Z}_{2^{l+s}}$. Then, the authenticator performs $\left \langle d \right \rangle \leftarrow \left \langle \phi  f(x) \right \rangle- \langle \phi \rangle \cdot \left \langle f(x) \right \rangle$. If $d = 0$, the verification succeeds; otherwise, abort.

\textbf{Function secret sharing} (FuncSS) \cite{boyle2015function,boyle2021function} within the 2PC setting divides a function $f$ into 2 succinct function shares $\{f_0,f_1\}$. Each party receives one of the function shares, which reveals no information about $f$.
For any input $x$, there exists $f_0(x)+f_1(x) = f(x)$. 
$\sysname$ leverages the comparison FuncSS $f^{<}_{a,b}$ \cite{boyle2021function,hao2023fastsecnet}, realized through distributed point functions (DPFs). Here, $f^{<}_{a,b}(x)$ evaluates to $b$ if $x<a$, and $0$ otherwise. Specifically, the comparison FuncSS $f^{<}_{a,b}$ consists of a pair of algorithms $\{\mathtt{Gen}^{<}(a,b), \mathtt{Eval}^{<}(\theta,\kappa_{\theta},x)\}$. The key generation algorithm $\mathtt{Gen}^{<}(a,b)$ outputs a pair of keys $\{\kappa_0, \kappa_1\}$, where each key implicitly represents $f_{\theta}^{<}$. For the party $\theta \in \{0,1\}$, with the key $\kappa_{\theta}$ and the public input $x$, the evaluation algorithm $\mathtt{Eval}^{<}(\theta,\kappa_{\theta},x)$ outputs $y_{\theta}$, i.e., the value of $f^{<}_{\theta}(x)$, where $f^{<}_{a,b}(x)= \sum_{\theta=0}^{1}y_{\theta}$.

\textbf{Multiple ideal functionalities} $\mathcal{F}_{\textnormal{ABB}}$ are used in $\sysname$ as arithmetic black-box operations, including $\mathcal{F}_{\textnormal{rand}}$, $\mathcal{F}_{\textnormal{coin}}$, $\mathcal{F}_{\textnormal{triple}}$, and others. These functionalities can be securely realized using well-established protocols in the malicious setting. Due to space constraints, we briefly outline their roles and refer to \cite{cramer2018spd,chida2023fast} for further details.

\begin{enumerate}[leftmargin=*,topsep=0pt] 
\item[-] $\mathcal{F}_{\textnormal{share}}(\mathbb{Z}_{2^{l+s}})$: Sample a pair of additive SS shares $\langle x \rangle  \stackrel{\$}{\leftarrow} \mathbb{Z}_{2^{{l+s}}}$ for the input $x$, such that $\langle x \rangle_0 +\langle x \rangle_1 \equiv_{l+s} x$.
\item[-] $\mathcal{F}_{\textnormal{rand}}(\mathbb{Z}_{2^{l+s}})$: Sample a random $r $, and share $\langle r\rangle \stackrel{\$}{\leftarrow} \mathbb{Z}_{2^{l+s}}$ between two parties.
\item[-] $\mathcal{F}_{\textnormal{coin}}(\mathbb{Z}_{2^{l+s}})$: Sample a random $r \stackrel{\$}{\leftarrow} \mathbb{Z}_{2^{l+s}}$, and output $r$ to two parties. 
\item[-] $\mathcal{F}_{\textnormal{triple}}(\mathbb{Z}_{2^{l+s}})$: Sample randoms $a,b$, compute $c=a\cdot b $, and output shared $\langle a\rangle_{\theta}, \langle b\rangle_{\theta}, \langle c\rangle_{\theta} \stackrel{\$}{\leftarrow} \mathbb{Z}_{2^{l+s}}$ to party $\theta$.
\item[-] $\mathcal{F}_{\textnormal{mult}}(\mathbb{Z}_{2^{l+s}})$: Take two additive shares $\langle x \rangle_\theta$ and $\langle y \rangle_\theta$ from party $\theta$, and output shared $\langle z\rangle_{\theta}\equiv_{l+s}\langle x \cdot y\rangle_{\theta}$ to party $\theta$.
\end{enumerate}

\section{System Overview}
\label{sec:ProblemStatement}

\begin{figure}[!t]
\centering
\includegraphics[width=0.48\textwidth]{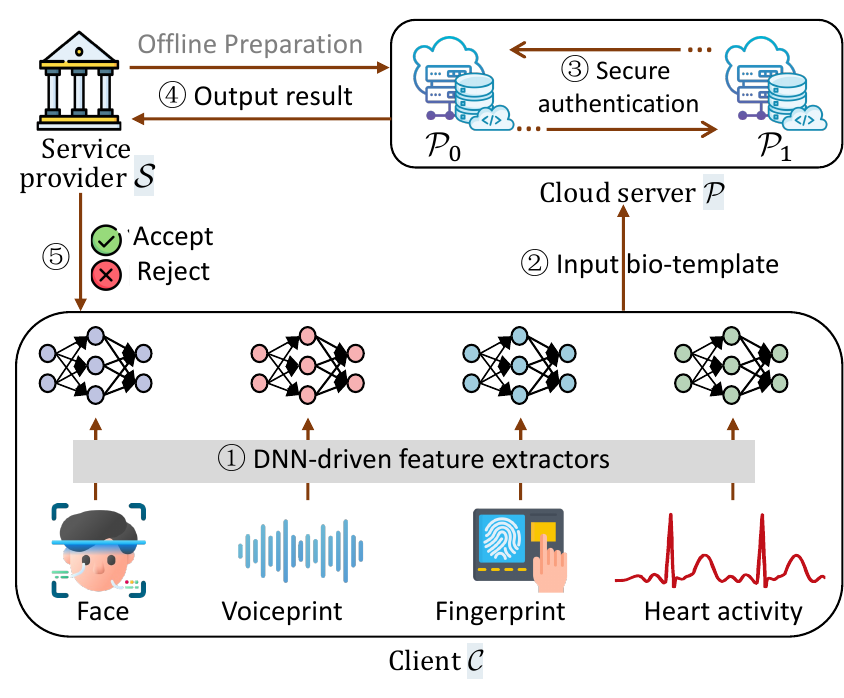}
\vspace{-6mm}
\caption{The system model.}
\vspace{-6mm}
\label{fig:SystemModel} 
\end{figure}

\textbf{System model.} Let us examine a scenario where a client requests access to a service, such as online tax declaration, using a biometric-based authentication system $\sysname$. 
This scenario is illustrated in Figure~\red{\ref{fig:SystemModel}}, consisting of three entities: the client $\mathcal{C}$, the service provider $\mathcal{S}$, and two cloud servers $\mathcal{P}_0$ and $\mathcal{P}_1$.  
Biometric authentication generally involves two phases: enrollment, where the client registers their reference biometric templates, and authentication, where the client is verified for subsequent logins. 
In $\sysname$, we assume that the enrollment phase has been completed and focus on the authentication phase.
Specifically, \circled{1} a client $\mathcal{C}$ generates a fresh biometric template via DNN-driven feature extractors and \circled{2} securely distributes it to $\mathcal{P}_{0}$ and $\mathcal{P}_{1}$. 
\circled{3} $\mathcal{P}_0$ and $\mathcal{P}_1$ perform secure authentication by computing the similarity between the $\mathcal{C}$'s fresh biometric template and the stored references.
They either identify the top-$k$ closest matches or determine whether the similarity score exceeds a predefined threshold $\tau$.
\circled{4} After authentication, $\mathcal{P}_0$ and $\mathcal{P}_1$ share the output with $\mathcal{S}$, cryptographically ensuring it is free from server-side tampering.
\circled{5} Upon receiving the authenticated decision result, $\mathcal{S}$ determines whether the client is granted access to the requested service.
In the two-server setting, each server retains the shares of biometric templates and intermediate results and no single server can unilaterally access the private information, thereby safeguarding privacy. This setting has been widely adopted in various secure applications \cite{cheng2024nomadic,wang2024cryptgraph,makri2021rabbit,wang2022towards}.

\textbf{Threat model and security.} There are two standard security models for outsourced computation: semi-honest security and malicious security. A semi-honest adversary follows the specifications of the protocols, but may attempt to infer additional information from the shares it handles. A malicious adversary may deviate arbitrarily from the protocols in an attempt to manipulate the computations. In $\sysname$, the two computing servers are assumed to be malicious, while the client and the service provider are considered semi-honest. 
Focusing specifically on the malicious security of the two-server computation (i.e., 2PC), we formally define the security of $\sysname$ within the real/ideal world.  
For simplicity, we assume that $\mathcal{P}_0$ is corrupted; the definition applies symmetrically to $\mathcal{P}_1$.

\begin{Definition}
\label{security}
Let $\mathcal{F}=\{\mathcal{F}_0, \mathcal{F}_1\}$ be an ideal-world functionality and $\prod$ be a real-world protocol semantical-securely computing $\mathcal{F}$.
For any malicious adversary $\mathcal{A}$, there exists a probabilistic polynomial-time (PPT) simulator $\mathsf{Sim}$ such that 
\begin{IEEEeqnarray}{rCL} 
&& \{\mathsf{View}^{\prod}_{\mathcal{A}(z)}(1^{l,s},x,y),\mathsf{Out}^{\prod}(1^{l,s},x, y)\}_{{l,s},z,x,y} \nonumber\\
& \overset{c}{\equiv }& \{\mathsf{Sim}^{\mathcal{F}}_{\mathcal{A}(z)}(1^{l,s},x,\mathcal{F}_0(x,y)), {\mathcal{F}}(x,y)\}_{l,s,z,x,y} \nonumber
\end{IEEEeqnarray}
where $l,s \in \mathbb{N}^{+}$ are the security parameters, $z\in\{0,1\}^{*}$ is the auxiliary information, $\mathsf{View}^{\prod}_{\mathcal{A}}$ denotes the final view of the corrupted $\mathcal{P}_0$ in the real world, $\mathsf{Sim}^{\mathcal{F}}$ denotes the view of the corrupted $\mathcal{P}_0$ generated by $\mathsf{Sim}$, $\mathsf{Out}^{\prod}$ denotes the output of the protocol $\prod$, ${\mathcal{F}}(x,y)$ denotes the output of the ideal functionality ${\mathcal{F}}$, and $ \overset{c}{\equiv }$ denotes computational indistinguishability against PPT adversaries except for a negligible advantage.
\end{Definition}

\textbf{Design goals.} Based on the adversarial model and system requirements, we establish the following design goals for $\sysname$:

\begin{itemize}[leftmargin=*,topsep=0pt] 
\item[-] \textbf{Privacy}. $\sysname$ must ensure that the client's biometric data, both fresh and pre-stored templates, remain confidential and are never exposed in plaintext to the servers and the service provider.  

\item[-] \textbf{Integrity}. $\sysname$ must guarantee that the authentication result is computed correctly and has not been tampered with, even if some servers behave maliciously. 
\item[-] \textbf{Efficiency}. Given the latency sensitivity of authentication tasks, $\sysname$ must minimize computational and communication overhead while maintaining strong security guarantees. 
\item[-] \textbf{Accuracy}. $\sysname$ must yield an authentication accuracy comparable to plaintext biometric systems, achieving a success rate no lower than $95\%$.
\item[-] \textbf{Scalability}. $\sysname$ should support large-scale deployments, efficiently handling many clients and high-dimensional biometric templates without significantly impacting performance. \end{itemize}

\section{Supporting Protocols}
\label{sec:SupportingProtocols}

This section leverages secret-sharing-family (SS-family) primitives to design linear inner product and the non-linear comparison protocols for $\sysname$. The high-level implementation of the functionality $\mathcal{F}_f$ is formally described in Figure~\ref{fig: functionality}. 
At a high level, $\sysname$ splits each protocol into offline and online phases, aiming to alleviate the computational burden and minimize the communication rounds in the online phase.
In the data-independent offline phase, heavy cryptographic operations---such as generating randomness, Beaver triples, and FuncSS keys---are executed with the assistance of a trusted dealer. In $\sysname$, the service provider $\mathcal{S}$ acts as the trusted dealer, distributing reliable correlated randomness to the two servers. 
For clarity, the main notations in this paper are summarized in Table~\ref{tab:Notations}.

\begin{table}[!t]
\centering
\caption{{The summary of notations}} 
\vspace{-3mm}
\label{tab:Notations}
\begin{adjustbox}{width=0.49\textwidth,center}
\begin{tabular}{l|p{9cm}}
\hline \hline  
\multicolumn{1}{l}{$\theta$} &  Party identity $\theta \in \{0,1\}$.    \\  
\multicolumn{1}{l}{$\mathcal{P}_0$, $\mathcal{P}_1$} &  Parties (i.e., Servers) running secure interactive computation.    \\  
\multicolumn{1}{l}{$\textbf{X}$, $X$, $x$ } &  Tensor/Matrix, vector, element.       \\  
\multicolumn{1}{l}{$ [1,x]$ }     & Set of positive integers $\{1, \dots, {x}\}$.   \\    
\multicolumn{1}{l}{$\langle x \rangle_{\theta  }$}    & Arithmetic share of $x$ held by party ${\theta  }$, where $x=\langle x \rangle_0+\langle x \rangle_1$.    \\ 
\multicolumn{1}{l}{$[\![x]\!]_{\theta}$}    & Optimized arithmetic share of $x$ held by party ${\theta}$, where $[\![x]\!]_{\theta}=(\Delta_{x}, \langle \lambda_{x} \rangle_{\theta})$ and $\Delta_{x}= x+ \lambda_{x}$.    \\  
\multicolumn{1}{l}{$X{[i]}$}         & The $i$-th element in vector $X$.        \\ 
\multicolumn{1}{l}{$\tau$}         & Authentication threshold.        \\ 
\multicolumn{1}{l}{$\phi$}         & Secret MAC key.        \\  
\multicolumn{1}{l}{$\otimes $} & Inner product operation for arithmetic values. \\ 
\multicolumn{1}{l}{$\cdot $} & Multiplication operation for arithmetic values.\\  \hline \hline
\end{tabular}
\end{adjustbox} 
\end{table}

\begin{figure}[!t]
\centering
\small
\fbox{
\parbox{0.9\linewidth}{
Linear function $f^{\text{lin}}$: $\mathbb{Z}_{2^l}\rightarrow \mathbb{Z}_{2^l}$. \\
\textbf{Input:}  $\mathcal{P}_{\theta}$ inputs $[\![\phi x]\!]_{\theta}$, $[\![x]\!]_{\theta}$, and $[\![ y]\!]_{\theta}$, $\forall \theta\in \{0,1\}$. \\
\textbf{Output:} $\mathcal{P}_{\theta}$ learns $[\![f^{\text{lin}}(x,y)]\!]_{\theta}$ and $[\![\phi f^{\text{lin}}(x,y)]\!]_{\theta}$, $\forall \theta\in \{0,1\}$.
\vspace{3mm} \\
Non-linear function $f^{\text{non-lin}}$: $\mathbb{Z}_{2^l}\rightarrow \mathbb{Z}_{2^l}$. \\
\textbf{Input:} $\mathcal{P}_{\theta}$ inputs $[\![x]\!]_{\theta}$ and $[\![\phi]\!]_{\theta}$, $\forall \theta\in \{0,1\}$. \\
\textbf{Output:} $\mathcal{P}_{\theta}$ learns $[\![f^{\text{non-lin}}(x)]\!]_{\theta}$ and $[\![\phi f^{\text{non-lin}}(x)]\!]_{\theta}$, $\forall \theta\in \{0,1\}$.
}}
\caption{Functionality $\mathcal{F}_f$.}
\label{fig: functionality}
\end{figure}

\subsection{Secure Inner Product Protocol}

\begin{figure}[!t] 
\small 
\centering
\begin{mybox}{Protocol $ \prod_{\mathtt{SecIP}}$} 
\textbf{\textcolor{gray}{\# Offline Phase:}} \\
\textbf{Input:} $\langle \lambda_{X}\rangle_{\theta }$, $\langle \lambda_{\phi X}\rangle_{\theta }$, $\langle \lambda_{Y}\rangle_{\theta }$. \\ 
\vspace{-3mm}
\begin{enumerate}[leftmargin=*,topsep=0pt] 
\item  Call $\mathcal{F}_\textnormal{rand}$ to learn the shared $\langle \lambda \rangle$-values: $\langle \lambda_{z} \rangle$ and $\langle \lambda_{\phi z} \rangle$.
\item  Call $\mathcal{F}_\textnormal{triple}$ to learn the shared $(\langle A_1\rangle,\langle B\rangle,\langle C_1 \rangle)$ and $(\langle A_2\rangle,\langle B\rangle,\langle C_2 \rangle)$, where $A_1\cdot B=C_1$ and $A_2\cdot B=C_2$.
\item $\forall \theta\in \{0,1\},$ $\mathcal{P}_{\theta}$ locally computes  $\langle \delta_X\rangle_{\theta} = \langle A_1\rangle_{\theta} -\langle \lambda_X\rangle_{\theta}$, $\langle \delta_{\phi X}\rangle_{\theta} = \langle A_2\rangle_{\theta} -\langle \lambda_{\phi X}\rangle_{\theta}$, and $\langle \delta_Y\rangle_{\theta} = \langle B\rangle_{\theta} -\langle \lambda_Y\rangle_{\theta}$.
\item Open $\langle \delta_X\rangle,\langle \delta_{\phi X}\rangle,\langle \delta_Y\rangle$ to get the clear $\delta_X, \delta_{\phi X},  \delta_Y$.
\end{enumerate} 
\textbf{Output:} $\mathcal{P}_{\theta}$ learns $\langle A_1\rangle_{\theta},\langle A_2\rangle_{\theta},\langle B\rangle_{\theta},\langle C_1\rangle_{\theta}, \langle C_2\rangle_{\theta}, \langle \lambda_{z}\rangle_{\theta},  \\\langle \lambda_{\phi z}\rangle_{\theta},
\delta_{\phi X},\delta_X,\delta_Y$. 
\vspace{3mm} \\
\textbf{\textcolor{gray}{\# Online Phase:}} \\
\textbf{Input:} For inputs $X$ and $Y$ with length $n$, $\mathcal{P}_{\theta}$ holds $(\Delta_{X}, \langle \lambda_X\rangle_{\theta }),(\Delta_{\phi X}, \langle \lambda_{\phi X}\rangle_{\theta })$, and $(\Delta_{Y}, \langle \lambda_Y\rangle_{\theta})$. \\ 
\vspace{-3mm}
\begin{enumerate}[leftmargin=*,topsep=0pt] 
\item $\mathcal{P}_{\theta}$ locally computes $\langle \Delta_{z} \rangle_{\theta}= \sum_{i=1}^{n}(\theta \cdot (\Delta_X[i]+\delta_X[i])\left(\Delta_Y[i]+\delta_Y[i]\right)-\langle A_1[i]\rangle \left(\Delta_Y[i]+\delta_Y[i]\right)-\left(\Delta_X[i]+\delta_X[i]\right)\langle B[i] \rangle+\langle C_1[i] \rangle) +\langle \lambda_{z}\rangle_{\theta}$.

\item $\mathcal{P}_{\theta}$ locally computes $\langle \Delta_{\phi z} \rangle_{\theta}=\sum_{i=1}^{n}(\theta \cdot (\Delta_{\phi X}[i]+\delta_{\phi X}[i])\left(\Delta_Y[i]+\delta_Y[i]\right)-\langle A_2[i]\rangle\left(\Delta_Y[i]+\delta_Y[i]\right)-\left(\Delta_{\phi X}[i]+\delta_{\phi X}[i]\right)\langle B[i] \rangle+\langle C_2[i] \rangle) +\langle \lambda_{\phi z}\rangle_{\theta}$. 

\item Open $\langle \Delta_{z} \rangle, \langle \Delta_{\phi z} \rangle$ to get the clear $\Delta_{z}, \Delta_{\phi z}$.

\item The servers run $\prod_{\textnormal{MACCheck}}$, if the check fails, abort.
\end{enumerate} 

\textbf{Output:} $\mathcal{P}_{\theta}$ learns $(\Delta_{z}, \langle  \lambda_{z}\rangle_{\theta})$ and $(\Delta_{\phi z}, \langle  \lambda_{\phi z}\rangle_{\theta})$.  
\end{mybox}
    \mycaptionspace
\caption{Construction of secure inner product protocol $\prod_{\mathtt{SecIP}}$.}  
\vspace{-3mm}
\label{alg:SecIP}
\end{figure} 

For $n$-dimensional OptSS shared $(\Delta_{X}, \langle \lambda_{X} \rangle)$, $\phi$-involved $(\Delta_{\phi X}, \langle \lambda_{\phi X} \rangle)$, and $(\Delta_{Y}, \langle \lambda_{Y} \rangle)$, $\langle \lambda \rangle$-related shares are independent of the actual inputs and stored in $\mathcal{P}_{0}$ and $\mathcal{P}_{1}$ in advance.
For an $n$-dimensional vector, the existing inner product computation has a complexity of $\mathcal{O}(n)$ over the secret-sharing domain. Inspired by \cite{yuan2024md}, we introduce an inner product protocol $\prod_{\mathtt{SecIP}}$ with online communication independent of the vector size $n$, while also optimizing the computation and communication overhead in the offline phase. 
Specifically, existing OptSS-based inner product protocols employ an open-then-sum strategy. This involves invoking $\prod_{\textnormal{mult}}$ (described in Sec.~\ref{sec:Preliminaries}) $n$ times, requiring the opening of $\langle  \Delta_Z[i] \rangle=\langle  Z[i]\rangle + \langle\lambda_Z[i]\rangle$ ($i \in \{1,n\}$) for each $\prod_{\textnormal{mult}}$ to retrieve $\Delta_Z[i]$, followed by summing up these values to learn $\Delta_z=\sum_{i=1}^{n}\Delta_Z[i]$. 
The key insight is that the open-then-sum strategy is equivalent to a sum-then-open strategy. In this optimized strategy, the intermediate shares $\langle  Z[i] \rangle$ are summed locally first with only a single random shared $\langle\lambda_z\rangle$, resulting in a single aggregated share $\langle  \Delta_z \rangle= \sum_{i=1}^{n}\langle  Z[i] \rangle+\langle\lambda_z\rangle$. $\langle  \Delta_z \rangle$ is then opened to retain $\Delta_z$. 
With this strategy, we observe that the MAC-related computations for $\Delta_{\phi z}$ do not need the extra full multiplication triples. Instead, these computations can share the part of multiplication triples (i.e., $B$) required for $\Delta_{z}$'s computation (refer to steps 2) and 3) of the offline phase in Figure.~\ref{alg:SecIP}). 
After completing the secure computation, two servers collaboratively verify the integrity of each revealed value with the MAC key $\phi$. Figure~\ref{alg:SecIP} presents the detailed online and offline phases for $\prod_{\mathtt{SecIP}}$.

To ensure data integrity, the MAC check protocol $\prod_{\text{MACCheck}}$ is used to detect discrepancies by verifying the consistency between shared data and its corresponding MAC-related values---any introduced errors will disrupt this consistency. To improve the efficiency of this verification, a batch MAC check protocol is developed (see Figure~\ref{MACCheck}), which has been adopted in prior works \cite{cramer2018spd, yuan2024md}. This protocol enables the simultaneous verification of multiple values in a constant number of rounds, significantly reducing communication overhead. Its compact and scalable design makes it particularly suitable for secure computation settings.

\begin{figure}[!t] 
\small 
\centering
\begin{mybox}{Protocol $\prod_{\textnormal{MACCheck}}$} 
On inputting a set of shared values $\{\langle X[i] \rangle\}_{i=1}^{n}$, shared MAC values $\{\langle \phi X[i]\rangle\}_{i=1}^{m}$, $\prod_{\textnormal{MACCheck}}$ executes with the inherent shared MAC key $\langle \phi \rangle$ and outputs True if the MAC check passes and abort otherwise.
\begin{itemize}[leftmargin=3pt,topsep=0pt] 
    \item Servers call $ \langle r\rangle\leftarrow \mathcal{F}_\textnormal{rand}(\mathbb{Z}_{2^s})$ and $\langle \phi r \rangle \leftarrow  \mathcal{F}_{\textnormal{mult}}(\langle \phi \rangle, 2^{l}\langle r \rangle)$. 
    \item  Servers call $ \mathcal{F}_\textnormal{coin}(\mathbb{Z}_{2^s})$ to sample public randomness $\{P[i]\}_{i=1}^{n}$. 
    \item  Servers locally compute $\langle y_0 \rangle \leftarrow 2^{l}\langle r \rangle +\sum_{i=1}^{n}{P[i] \cdot \langle X[i] \rangle}$ and $\langle y_1 \rangle \leftarrow \langle \phi r \rangle +\sum_{i=1}^{n}P[i]\langle \phi {X[i]}\rangle \pmod{2^{l+s}}$.
    \item Each server broadcasts $\langle y_0 \rangle \in \mathbb{Z}^{l+s}$ and all servers compute $y_0= \langle y_0 \rangle_0+\langle y_0 \rangle_1$.
    \item  Servers locally compute $\langle z\rangle \leftarrow \langle y_1 \rangle- y_0 \langle \phi \rangle$, then open their commitments.
    \item Servers verify $\langle z\rangle_0+\langle z\rangle_1 \equiv_{l+s} 0$. If it outputs True then all servers proceed and abort otherwise.
    \end{itemize}
\end{mybox}
    \mycaptionspace
\caption{The batch MAC check protocol. 
\label{MACCheck} } 
\mycaptionspace
\end{figure}

\begin{theorem}[The correctness of $\prod_{\mathtt{SecIP}}$]
\label{theo1}
For any OptSS-shared input vectors $(\Delta_X,\left<\lambda_X\right>)$, $(\Delta_{\phi X},\left<\lambda_{\phi X}\right>)$, and $(\Delta_Y,\left<\lambda_Y\right>)$, the $\prod_{\mathtt{SecIP}}$ yields the correct results $(\Delta_z,\left<\lambda_z\right>) $ and $(\Delta_{\phi z},\left<\lambda_{\phi z}\right>) $ if the conditions $\Delta_z - \lambda_z=z= X\otimes Y=\sum_{i=1}^{n}X[i]\cdot Y[i]$ and $\Delta_{\phi z} - \lambda_{\phi z}=\phi z= \phi X\otimes Y=\sum_{i=1}^{n}\phi X[i]\cdot Y[i]$ are satisfied, with $\otimes$ represents the inner product operation.
\end{theorem}

\begin{proof} 
$\prod_{\mathtt{SecIP}}$ consists of offline and online phases described in Figure~\ref{alg:SecIP}. The correctness of the offline phase is guaranteed through two main components. First, the functionalities $\mathcal{F}_\textnormal{rand}$ and $\mathcal{F}_\textnormal{triple}$ reliably generate the required $\lambda$-values and multiplication triples, guaranteeing their correctness. 
Second, the $\delta$-values, which prepare inputs for the online phase, are computed locally using the relations $\delta_X = A_1 -\lambda_X$, $\delta_{\phi X} = A_2 -\lambda_{\phi X}$, and $\delta_Y = B -\lambda_Y$, ensuring proper input adjustments for subsequent evaluation.  

Next, we prove the correctness of the online phase of the $\prod_{\mathtt{SecIP}}$ protocol. Before producing the shared value $\langle z \rangle = \langle X \otimes Y \rangle$, we first establish the correctness of $\langle X[i] \cdot Y[i] \rangle$ for each $i$, as outlined below:

\vspace{-2mm} 
\begin{small}
\begin{IEEEeqnarray}{rCL}
&&\langle Z[i] \rangle_0 + \langle Z[i] \rangle_1 \nonumber \\
&=&\left(\Delta_X[i]+\delta_X[i]\right)\left(\Delta_Y[i]+\delta_Y[i]\right) -\langle A[i]\rangle_1 \left(\Delta_Y[i]+\delta_Y[i]\right)\nonumber \\
&&- \left(\Delta_X[i]+\delta_X[i]\right)\langle B[i] \rangle_1 +\langle C[i]\rangle_1-\langle A[i]\rangle_0 \left(\Delta_Y[i]+\delta_Y[i]\right)\nonumber \\
&&- \left(\Delta_X[i]+\delta_X[i]\right)\langle B[i] \rangle_0+\langle C[i] \rangle_0 \nonumber  \\
&&\textcolor{gray}{\# \small{\Delta_X[i]+\delta_X[i] =X[i] +A[i]\leftarrow X[i] +\lambda_X[i]+A[i]-\lambda_X[i]}} \nonumber  \\ 
&&\textcolor{gray}{\#\small{\Delta_Y[i]+\delta_Y[i] =Y[i] +B[i] \leftarrow Y[i] +\lambda_Y[i]+B[i]-\lambda_Y[i]}} \nonumber \\
&=& (X[i] +A[i])(Y[i] +B[i] )- A[i](Y[i] +B[i] )\nonumber \\
&&-(X[i] +A[i])B[i] +C[i]\nonumber \\
&=&X[i] \cdot Y[i].  \nonumber
\end{IEEEeqnarray}   
\end{small}

\noindent By summing over all $i \in [1, n]$, the protocol ensures that $\langle z \rangle = \sum_{i=1}^{n} \langle X[i] \cdot Y[i]\rangle $ is correct. Consequently, $\Delta_z =z +\lambda_z$ is proven correct by opening $\langle \Delta_z \rangle$, where $\langle \Delta_z \rangle= \langle z\rangle+\langle \lambda_{z}\rangle \leftarrow\sum_{i=1}^{n} \langle X[i] \cdot Y[i] \rangle +\langle \lambda_{z}\rangle$. The correctness of $\Delta_{\phi z} =\phi X \otimes Y +\lambda_{\phi z}$ is established in the same manner.
\end{proof}

\subsection{Secure Comparison Protocol}
\begin{figure}[!t] 
\small 
\centering
\begin{mybox}{Protocol $ \prod_{\mathtt{SecCMP}}$}  
\textbf{\textcolor{gray}{\# Offline Phase:}} \\
\textbf{Input:} $\langle  \phi  \rangle_{\theta }$ and $\langle \lambda_{x}\rangle_{\theta }$ . \\ 
\vspace{-3mm}
\begin{enumerate}[leftmargin=*,topsep=0pt]  
\item Let $\textbf{b}=(b_0,b_1)=(1,\phi)$ and $a=\lambda_x$.
\item $\left(\kappa_0^{\prime}, \kappa_1^{\prime}\right) \leftarrow \mathtt{Gen}_{a,-\textbf{b}}^{<}$.
\item Call $\mathcal{F}_\textnormal{share}$ to share $b_0$ and $b_1$ to learn the shared $\langle b_0 \rangle,\langle b_1 \rangle \leftarrow \mathbb{Z}_{2^{l+s}}$, s.t., $\langle b_0\rangle_0+\langle b_0 \rangle_1\equiv_{l+s}b_0 $, and $\langle b_1\rangle_0+\langle b_1\rangle_1\equiv_{l+s} b_1 $.
\item Let $\kappa_{\theta }=\kappa_{\theta }^{\prime} ||\langle \textbf{b} \rangle_{\theta }$ for $\theta \in\{0,1\}$.
\item  Call $\mathcal{F}_\textnormal{rand}$ to learn the shared $\langle \lambda_z \rangle$, $\langle \lambda_{\phi z} \rangle$.
\end{enumerate} 
\textbf{Output:} $\mathcal{P}_{\theta}$ learns $\kappa_{\theta}$, $\langle \lambda_z \rangle_{\theta}$, $\langle \lambda_{\phi z} \rangle_{\theta}$.
\vspace{3mm} \\
\textbf{\textcolor{gray}{\# Online Phase:}} \\
\textbf{Input:} For any inputs $x$, $\mathcal{P}_{\theta}$ holds 
$(\Delta_{x}, \langle \lambda_x\rangle_{\theta })$. \\ 
\vspace{-3mm}
\begin{enumerate}[leftmargin=*,topsep=0pt] 
\item Parse $\kappa_{\theta }=\kappa_{\theta }^{\prime}||\langle \textbf{b} \rangle_{\theta }$. 
\item Set $(\langle  \gamma_0 \rangle_{\theta }, \langle  \gamma_1 \rangle_{\theta }) \leftarrow \mathtt{Eval}_{a,-\textbf{b}}^{<}\left(\theta, \kappa_{\theta }^{\prime}, \Delta_x \right)$.
\item Set $\langle  \Delta_{z} \rangle_{\theta } \leftarrow \langle  \gamma_0 \rangle_{\theta }+ \langle b_0 \rangle_{\theta } + \langle \lambda_z  \rangle_{\theta}$.
\item Set $ \langle  \Delta_{\phi z} \rangle_{\theta } \leftarrow \langle  \gamma_1 \rangle_{\theta }+ \langle b_1 \rangle_{\theta } + \langle \lambda_{\phi z}  \rangle_{\theta}$.
\item Open $\langle \Delta_{z} \rangle, \langle \Delta_{\phi z} \rangle$ to get the clear $\Delta_{z}, \Delta_{\phi z}$.
\item The servers run $\prod_{\textnormal{MACCheck}}$, if the check fails, abort.
\end{enumerate} 
\textbf{Output:} $\mathcal{P}_{\theta}$ learns $(\Delta_{z}, \langle \lambda_z \rangle_{\theta})$, $(\Delta_{\phi z},\langle \lambda_{\phi z} \rangle_{\theta})$.
\end{mybox}
    \mycaptionspace
\caption{Construction of secure comparison protocol $ \prod_{\mathtt{SecCMP}}$.}  
\vspace{-3mm}
\label{fig:SecCMP}
\end{figure} 

The FuncSS-based comparison operation presented in Sec.~\ref{sec:Preliminaries} operates with $f_{a,b}^{<}(x)$, designed for less-than evaluations where $x<a\Rightarrow b$. However, there are two challenges in employing $f_{a,b}^{<}(x)$ for $\sysname$. 
\textit{Challenge 1}: The generic $\sysname$ requires functionality for greater-than comparisons, specifically to assess whether a similarity score $ \mathfrak{s}$ exceeds a predefined authentication threshold $\tau$ or retrieve the top-$k$ most similar matches. This can be expressed as $(\mathfrak{s}\geq \tau)?b:0$, where $b$ represents the output when the condition is satisfied.
\textit{Challenge 2}: Extending this non-verifiable $f_{a,b}^{<}(x)$ to a lightweight, verifiable 2PC-FuncSS construction poses another challenge. 
To address these issues, we make two key technical observations. 

\begin{itemize}[leftmargin=*,topsep=0pt]
\item[-] \textit{For challenge 1}: We employ the observation that $f_{a, b}^{\geq}(x)$= $b+f_{a,-b}^{<}(x)$ where $x\geq a $ $\Rightarrow$ $ b$. This formulation reuses the less-than comparison functionality, thereby eliminating the need for introducing new crypto-primitives or altering the protocol's core structure. This transformation incurs no additional overhead in terms of computation or communication. The additive operation $+$ within the secret-sharing domain is inherently \textit{free}, as it involves purely local computation without any inter-party interaction. 
\item[-] \textit{For challenge 2}: The shares of $\phi$-related results (i.e., $\phi f_{a, b}^{\geq}(x)$) are either $0$ or $ \phi b$, depending on whether $f_{a, b}^{\geq}(x)$ is $0$ or $b$. Inspired by \cite{boyle2021function,hao2023fastsecnet}, we employ $f_{a, b}^{\geq}(x)$ to output a pair of authenticated coefficients. More concretely, the construction outputs the coefficients $\textbf{b}=(b_0,b_1)=(b,\phi b)$ for $f_{a, \textbf{b}}^{\geq}(x)$ when $x \geq a$, otherwise $(0,0)$.
\end{itemize}
Notably, in $\sysname$, the value of $b$ is fixed to $1$, i.e., $\textbf{b}=(1,\phi)$.
Building on these two observations, we propose a verifiable greater-than comparison protocol $\prod_{\mathtt{SecCMP}}$. The detailed procedures for both the online and offline phases of $\prod_{\mathtt{SecCMP}}$ are illustrated in Figure~\ref{fig:SecCMP}.

\begin{theorem}[The correctness of $\prod_{\mathtt{SecCMP}}$]
\label{theo2}
Given any OptSS-shared input $(\Delta_x, \langle \lambda_x \rangle)$, $\prod_{\mathtt{SecCMP}}$ yields the correct results $(\Delta_z,\left<\lambda_z\right>) $ and $(\Delta_{\phi z},\left<\lambda_{\phi z}\right>)  $ if the conditions $\Delta_z - \lambda_z = 1\{x \geq 0\}$ and $\Delta_{\phi z} - \lambda_{\phi z}= \phi\{x \geq 0\}$ are satisfied.
\end{theorem}
\begin{proof} 
Likewise, $\prod_{\mathtt{SecCMP}}$ consists of offline and online phases described in Figure~\ref{fig:SecCMP}. The offline phase involves invoking the key generation algorithm $\mathtt{Gen}^{<}$ to produce a pair of keys $\{{\kappa}_0^{\prime}, {\kappa}_1^{\prime}\}$, $\mathcal{F}_\textnormal{share}$ to generate related additive shares, and $\mathcal{F}_\textnormal{rand}$ to reliably generate the required $\lambda$-values. The correctness of $\mathtt{Gen}^{<}$ and $\mathtt{Eval}^{<}$ has been directly proven in \cite{boyle2021function}.
Now, let's prove the correctness of the online phase of $\prod_{\mathtt{SecCMP}}$ in producing $\Delta_z, \langle \lambda_z \rangle$, as outlined below:
%

\begin{IEEEeqnarray}{rCL}
&&\langle \Delta_z \rangle_0+\langle \Delta_z \rangle_1  \nonumber  \\ 
&=&   \mathtt{Eval}_{\lambda_x,-1}^{<}(0, {\kappa}_{0}^{\prime}, \Delta_x)_{0}  + \langle 1\rangle_0 +\langle \lambda_z \rangle_0 \nonumber \\
&&+\mathtt{Eval}_{\lambda_x,-1}^{<}(1, {\kappa}_{1}^{\prime}, \Delta_x)_{1}  +\langle 1\rangle_1 +\langle \lambda_z \rangle_1 \nonumber  \\ 
&=& -1\left\{ x+\lambda_x <\lambda_x  \right\} + 1+ \lambda_z  = 1\left\{ x\geq 0\right\}+ \lambda_z  \nonumber \\ 
&\Rightarrow& \Delta_z - \lambda_z =z = 1\left\{ x\geq 0\right\}. \nonumber  
\end{IEEEeqnarray}

\vspace{-4mm}

\begin{IEEEeqnarray}{rCL}
&&\langle \Delta_{\phi z} \rangle_0+\langle \Delta_{\phi z} \rangle_1  \nonumber  \\ 
&=&   \mathtt{Eval}_{\lambda_x,-\phi}^{<}(0, {\kappa}_{0}^{\prime}, \Delta_x)_{0}  + \langle \phi \rangle_0 +\langle \lambda_{\phi z} \rangle_0 \nonumber \\
&&+\mathtt{Eval}_{\lambda_x,-\phi}^{<}(1, {\kappa}_{1}^{\prime}, \Delta_x)_{1}  +\langle \phi \rangle_1 +\langle \lambda_{\phi z} \rangle_1 \nonumber  \\ 
&=& -\phi\left\{ x+\lambda_x <\lambda_x  \right\} + \phi + \lambda_{\phi z}  = \phi \left\{ x\geq 0\right\}+ \lambda_{\phi z}  \nonumber \\ 
&\Rightarrow& \Delta_{\phi z} - \lambda_{\phi z} = {\phi z} = \phi \left\{ x\geq 0\right\}. \nonumber  \hfill  \qedhere 
\end{IEEEeqnarray}
\end{proof}

\section{The Design of \sysname}
\label{sec:SystemDesign} 

Building on the components introduced in Sec.~\ref{sec:SupportingProtocols}, 
we construct the $\sysname$ scheme. Conceptually, $\sysname$ comprises two key phases: client-side biometric template pre-processing and outsourcing and server-side secure authentication. Consistency check $\prod_{\textnormal{MACCheck}}$ is executed in each component of the secure authentication phase to ensure correctness against malicious adversaries.

\subsection{Client-side Pre-processing and Outsourcing}

Existing PPBA studies primarily utilize the Euclidean distance or the Cosine distance to measure similarity. Below, we formally define these two metrics for two vectors $X = [X[i]]_{i=1}^{n}$ and $Y= [Y[i]]_{i=1}^{n}$. The Cosine distance between $X $ and $Y $ is formulated as
\begin{IEEEeqnarray}{rCL}
 \text{CosD}(X,Y) = \frac{\sum_{i=1}^n X[i]\cdot Y[i]}{\sqrt{\sum_{i=1}^n X[i]^2} \cdot \sqrt{\sum_{i=1}^n Y[i]^2}}.  
\label{CosS}
\end{IEEEeqnarray}
\noindent The Euclidean distance is defined as $\text{EucD}(X,Y) = {\sum_{i=1}^{n} (X[i] - Y[i])^2}$.
From Eq.~\ref{CosS}, it is evident that computing Cosine distance inherently involves expensive non-linear operations, namely division and square root. To mitigate this issue, Cheng \etal~\cite{cheng2024nomadic} proposed an MPC-friendly reformulation by integrating similarity score computation with authentication matching. Specifically, the authentication condition $1\{\text{CosD}(X,Y) \geq \tau\}$ (where $\tau$ is the public authentication threshold in plaintext) is transformed into $1\{\text{sign}(X \otimes Y) \wedge \text{sign}(\frac{1}{\tau^2} (X \otimes Y)^2 $-$ (X \otimes X)(Y \otimes Y))\}$,
where $\wedge$ denotes the logical AND operation, and $\text{sign}(x)$ is equivalent to the comparison operation that outputs 1 if $x \geq 0$ and 0 otherwise.
This transformation effectively eliminates division and square root operations but introduces \textit{two comparison and three inner product operations} per authentication. In contrast, Euclidean distance-based authentication merely involves \textit{one comparison and one inner product operations}. 

$\sysname$ aims to enable service-selectable similarity metrics while maintaining server-side compatibility. That means a unified authentication procedure on the server side applies regardless of whether Euclidean or Cosine distance is used. However, this presents two key challenges:
\begin{itemize}[leftmargin=*,topsep=0pt]
    \item[-] \textit{Challenge 1:} Eliminating division and square root operations from Cosine distance to align with Euclidean distance computation.
    \item[-] \textit{Challenge 2:} Bridging the semantic gap between the two metrics—Cosine distance increases with similarity, whereas Euclidean distance decreases with similarity.
\end{itemize}

To address these challenges and achieve an efficient yet flexible authentication procedure, the client pre-processes the DNN-extracted biometric template $T=\left [T[1],T[2],\cdots,T[n]\right]$ according to the similarity measure $\mathfrak{m}\in \{0,1\}$. Here, $\mathfrak{m}=0$ corresponds to Cosine distance, and Euclidean distance otherwise. 
For $\mathfrak{m}=1$, the biometric template $T$ is normalized to a unit-length vector, yielding $\widehat{T}=\left [\frac{T[1]}{||T||},\frac{T[2]}{||T||},\cdots,\frac{T[n]}{||T||}, 0\right]$, where the $L^2$ norm $||T||=  \sqrt{\sum_{i=1}^{n}T[i]^2}$. Since the transformation guarantees $||\widehat{T}|| =\sqrt{\sum_{i=1}^{n}\widehat{T}[i]^2}=1$, explicit division and square root computations can be omitted during similarity computations (resolving \textbf{Challenge 1}).
To reduce the impact of pre-processing on authentication accuracy while achieving storage compression, we adopt the method proposed in \cite{engelsma2021learning}, which compresses each floating-point vector element into an 8-bit integer using min-max normalization.
%
Conversely, for $\mathfrak{m}=1$, the biometric template $T$ is expanded by introducing an extra dimension, resulting in $\widehat{T}=\left [T[1],T[2],\cdots,T[n], -\frac{1}{2} \left(\sum_{i=1}^{n}T[i]^2 \right)\right]$ for registration phase and $\widehat{T}=\left [T[1],T[2],\cdots,T[n], 1\right]$ for authentication phase.
This expansion bridges the semantic gap between the two metrics, making the equivalent Euclidean distance (i.e., computed by the inner product directly) increase with similarity (resolving \textbf{Challenge 2}).
The correctness is demonstrated below.  

Given the processed fresh biometric template $\widehat{T}$, and two stored reference biometric templates $\widehat{{D}}_1$ and $\widehat{{D}}_2$, their respective inner products with $\widehat{T}$ are computed as: 
\begin{IEEEeqnarray}{rCL}
\widehat{T} \otimes \widehat{{D}}_j&=&\sum\nolimits_{i=1}^{n+1} \widehat{T}[i] \cdot  \widehat{{D}}_j[i] \nonumber \\
&=&\sum\nolimits_{i=1}^{n}\widehat{T}[i] \cdot  \widehat{{D}}_j[i]  - \frac{1}{2}\sum\nolimits_{i=1}^{n}\widehat{{D}}_j^{2}[i], \forall j \in \{1,2\}. \nonumber  
\end{IEEEeqnarray}
The squared Euclidean distance difference between $\widehat{T}$ and the two stored templates simplifies to:

\begin{small}
\begin{IEEEeqnarray}{rCL}
&&\text{EucD}(\widehat{T},\widehat{{D}}_1) - \text{EucD}(\widehat{T},\widehat{{D}}_2) \nonumber \\
&=& \sum_{i=1}^{n}\widehat{{D}}_1[i]^2 - 2\sum_{i=1}^{n}\widehat{T}[i]\cdot \widehat{{D}}_1[i]-\sum_{i=1}^{n}\widehat{{D}}_2[i]^2+2\sum_{i=1}^{n}\widehat{T}[i]\cdot \widehat{{D}}_2[i]
\nonumber \\
&=& 2(\sum_{i=1}^{n}\widehat{T}[i]\cdot \widehat{{D}}_2[i]-\frac{1}{2}\widehat{{D}}_2[i]^2)- 2(\sum_{i=1}^{n}\widehat{T}[i]\cdot \widehat{{D}}_1[i]-\frac{1}{2}\widehat{{D}}_1[i]^2)\nonumber \\
&=&2(\widehat{T} \otimes \widehat{{D}}_2-\widehat{T} \otimes \widehat{{D}}_1). \nonumber
\end{IEEEeqnarray}
\end{small}

\noindent Thus, $\text{EucD}(\widehat{T},\widehat{{D}}_1) \leq \text{EucD}(\widehat{T},\widehat{{D}}_2) \iff  \widehat{T} \otimes \widehat{{D}}_1 \geq \widehat{T} \otimes \widehat{{D}}_2$ is learned. This confirms that the inner product serves as an inverse surrogate for the Euclidean distance. 
This pre-processing guarantees a unified authentication procedure on the server side for different similarity metrics.

$\mathcal{C}$ then distributes the processed template $\widehat{T} =\left [T[1],T[2],\cdots,T[n],T[n+1]\right]$ between the two servers $\mathcal{P}_0$ and $\mathcal{P}_1$ with arithmetic secret sharing. Specifically, $\mathcal{C}$ selects a random vector ${R}\in \mathbb{Z}_{2^{l+s}}^{n+1}$, sending $\langle \widehat{T} \rangle_0 = {R}$ to $\mathcal{P}_0$ and $\langle \widehat{T}\rangle_1 = \widehat{T}-{R}$ to $\mathcal{P}_1$. This entire dimensionality extension process is completed within a few microseconds 
and its cost is effectively amortized over subsequent authentication.

\begin{figure}[!t] 
\small 
\centering
\begin{mybox}{Protocol $ \prod_{\text{Initial}}$} 
\label{ArithMult} 
\textbf{Input:} For any inputs $\widehat{\textbf{D}}$ with identity index $I$ and $\phi$, $\mathcal{P}_{\theta}$ hold $\langle \widehat{\textbf{D}}\rangle_{\theta}$ and $\langle \phi \rangle_{\theta}$, here $\langle \vec{\textbf{D}}\rangle_{\theta}=[\langle I \rangle_{\theta}, \langle \widehat{\textbf{D}}\rangle_{\theta}]$. \\
\begin{enumerate}[leftmargin=*,topsep=0pt]  
\item  Call $\mathcal{F}_\textnormal{rand}$ to learn the shared $\langle \lambda_{ \vec{\textbf{D}}} \rangle$, $\langle \lambda_{\phi} \rangle$, $\langle \lambda_{\phi \vec{\textbf{D}}} \rangle$, $\langle \lambda_{\widehat{T}} \rangle$.
\item  Call $\mathcal{F}_\textnormal{triple}$ to learn the shared multiplication triples $(\langle \textbf{A}\rangle,\langle \textbf{B} \rangle,\langle \textbf{C} \rangle)$.
\item Call $\prod_\textnormal{OptSS}(\langle \vec{\textbf{D}} \rangle)$ and $\prod_\textnormal{OptSS}(\langle  \phi \rangle)$ to learn {$(\Delta_{\vec{\textbf{D}}}, \langle \lambda_{\vec{\textbf{D}}}\rangle_{\theta })$} and $(\Delta_{\phi},\langle \lambda_{\phi}\rangle_{\theta})$.
\item $\forall \theta\in \{0,1\},$ $\mathcal{P}_{\theta}$ locally computes $\langle \delta_\phi\rangle_{\theta} = \langle \textbf{A}\rangle_{\theta} -\langle \lambda_\phi\rangle_{\theta}$ and $\langle \delta_{\vec{\textbf{D}}}\rangle_{\theta} = \langle \textbf{B}\rangle_{\theta} -\langle \lambda_{\vec{\textbf{D}}}\rangle_{\theta}$.
\item Open $\langle \delta_\phi\rangle, \langle \delta_{\vec{\textbf{D}}}\rangle$ to get the clear $\delta_\phi, \delta_{\vec{\textbf{D}}}$. 
\item $\mathcal{P}_{\theta}$ locally computes $\langle \Delta_{\phi  \vec{\textbf{D}}} \rangle= (\Delta_\phi+\delta_\phi)\left(\Delta_{\vec{\textbf{D}}}+\delta_{\vec{\textbf{D}}}\right)-\langle \textbf{A}\rangle \left(\Delta_{\vec{\textbf{D}}}+\delta_{\vec{\textbf{D}}}\right)-\left(\Delta_\phi+\delta_\phi\right)\langle \textbf{B} \rangle+\langle \textbf{C} \rangle +\langle \lambda_{\phi  \vec{\textbf{D}}}\rangle$.
\item Open $\langle \Delta_{\phi  \vec{\textbf{D}}} \rangle$ to get the clear $ \Delta_{\phi  \vec{\textbf{D}}}=\phi \vec{\textbf{D}} +\lambda_{\phi \vec{\textbf{D}}}$. 
\end{enumerate} 
\textbf{Output:} $\mathcal{P}_{\theta}$ splits {$(\Delta_{\vec{\textbf{D}}}, \langle \lambda_{\vec{\textbf{D}}}\rangle_{\theta })$} and {$ (\Delta_{\phi  \vec{\textbf{D}}}, \langle \lambda_{\phi \vec{\textbf{D}}}\rangle_{\theta})$} to learn $ (\Delta_{ \widehat{\textbf{D}}}, \langle \lambda_{ \widehat{\textbf{D}}}\rangle_{\theta})$ and $ (\Delta_{I}, \langle \lambda_{I}\rangle_{\theta})$ and $ (\Delta_{\phi  \widehat{\textbf{D}}}, \langle \lambda_{\phi \widehat{\textbf{D}}}\rangle_{\theta})$ and $ (\Delta_{\phi  I}, \langle \lambda_{\phi I}\rangle_{\theta})$, as well as $(\Delta_{\phi},\langle \lambda_{\phi}\rangle_{\theta})$, and $\langle \lambda_{\widehat{T}} \rangle_{\theta}$.
\end{mybox}
    \mycaptionspace
\caption{The initialization of our $\sysname$. \label{fig:initial}} 
\vspace{-3mm}
\end{figure} 

\begin{algorithm}[!t]
    \caption{Secure Authentication Phase in $\sysname$. \label{alg:Scheme}}
    \begin{algorithmic}[1] 
    \Require $\mathcal{P}_0$ and $\mathcal{P}_1$ receive shared $[\langle\widehat{T}\rangle,\langle I_{\mathcal{C}}\rangle]$. 
    \Ensure $\mathcal{P}_0$ and $\mathcal{P}_1$ output $(\Delta_{\eta},\langle \lambda_{\eta}\rangle_{\theta})$ to $\mathcal{S}$.
    \item Call $\prod_\textnormal{OptSS}([\langle\widehat{T}\rangle,\langle I_{\mathcal{C}}\rangle] )$ to learn $[\![\widehat{T}]\!]=(\Delta_{\widehat{T}},\langle \lambda_{\widehat{T}}\rangle)$ and $[\![I_{\mathcal{C}}]\!]=(\Delta_{I_{\mathcal{C}}},\langle \lambda_{I_{\mathcal{C}}}\rangle)$.
    \State  \textbf{\textcolor{gray}{\# Similarity score computation:}} 
    \For{$i = 1 \to m$}
        \State $[\![S[i]]\!], [\![\phi S[i]]\!]=\mathtt{SecIP}([\![\widehat{\textbf{D}}[i]]\!], [\![\phi \widehat{\textbf{D}}[i]]\!], [\![\widehat{T}]\!])$.  
    \EndFor   
    \State \textbf{\textcolor{gray}{\# Top-1 similarity retrieval:}} 
    \State $[\![\tilde{S}]\!]  = [\![S[1]]\!]$, $[\![\tilde{I}]\!]  = [\![I[1]]\!]$. 
    \State $[\![\tilde{\phi S}]\!]  = [\![\phi S[1]]\!]$, $[\![\tilde{\phi I}]\!]  = [\![\phi I[1]]\!]$. 
    \For{$i = 2 \to m$}
        \State $ [\![\mathfrak{b}]\!], [\![\phi \mathfrak{b}]\!]= \mathtt{SecCMP}([\![\tilde{S}]\!]  - [\![S[i]]\!])$.
        \State $[\![\tilde{S}]\!] = [\![S[i]]\!]+\prod_{\textnormal{mult}}{([\![\mathfrak{b}]\!], [\![\tilde{S}]\!]-[\![S[i]]\!])}$ and $[\![\tilde{\phi S}]\!] = [\![\phi S[i]]\!]+\prod_{\textnormal{mult}}{([\![\mathfrak{b}]\!], [\![\tilde{\phi S}]\!]-[\![\phi S[i]]\!])}$.
        \State $[\![\tilde{I}]\!] = [\![I[i]]\!]+\prod_{\textnormal{mult}}{([\![\mathfrak{b}]\!], [\![\tilde{I}]\!]-[\![I[i]]\!])}$ and $[\![\tilde{\phi I}]\!] = [\![\phi I[i]]\!]+\prod_{\textnormal{mult}}{([\![\mathfrak{b}]\!], [\![\tilde{\phi I}]\!]-[\![\phi I[i]]\!])}$.
    \EndFor 
    \State  \textbf{\textcolor{gray}{\# Identity Match:}} 
    \State $[\![\eta]\!] = [\![\tilde{I}]\!] - [\![I_{\mathcal{C}}]\!]$ and $[\![\phi \eta]\!] = [\![\phi \tilde{I}]\!] - [\![\phi ]\!] \cdot [\![I_{\mathcal{C}}]\!]$.
    \item The servers run $\prod_{\textnormal{MACCheck}}$ to check the each secure computation result. If the check fails, abort.
    \State Return $[\![\eta]\!]$ to the service provider for further decision.
    \end{algorithmic}
\end{algorithm}

\subsection{Server-side Secure Authentication} In $\sysname$, we assume $m$ clients have already completed registration, and the servers have pre-stored a large number of shared reference biometric templates $\langle \widehat{\textbf{D}}\rangle$ with identity index $\langle I\rangle$ ($ \forall I\in [1,m ]$), that is $\langle \vec{\textbf{D}}\rangle_{\theta}=[\langle I \rangle_{\theta}, \langle \widehat{\textbf{D}}\rangle_{\theta}]$. 
Given the MAC key $\langle \phi \rangle$ from the trusted $\mathcal{S}$, the two servers $\mathcal{P}_0$ and $\mathcal{P}_1$ initialize to compute $(\Delta_{\vec{\textbf{D}}}, \langle\lambda_{{\vec{\textbf{D}}}}\rangle)$ and $(\Delta_{\phi \vec{\textbf{D}}}, \rangle\lambda_{ {\phi \vec{\textbf{D}}}}\rangle)$, as shown in Figure~\ref{fig:initial}.
Upon receiving the shared fresh biometric template $\langle \widehat{T}\rangle$ from the client $\mathcal{C}$ (i.e., the identity index is $I_{\mathcal{C}}$), $\mathcal{P}_0$ and $\mathcal{P}_1$ jointly execute the authentication phase, shown in Algorithm~\ref{alg:Scheme}. In this phase, the servers securely compute the similarity scores between \( \langle \widehat{T} \rangle \) and the stored biometric database $\langle \widehat{\textbf{D}}\rangle$ without revealing any sensitive information. The computation leverages $\prod_{\mathtt{SecIP}}$ to obtain a similarity score vector $[\![S[i]]\!], [\![\phi S[i]]\!]$, where each entry corresponds to the similarity between \( \langle \widehat{T} \rangle \) and the $i$-th stored biometric template. Finally, the servers collaboratively determine the most similar entry (i.e., top-1 nearest neighbor) using a secure maximum-finding protocol, yielding the shared authentication result $[\![\eta]\!], [\![\phi \eta]\!]$. 
At the end of the online secure authentication phase, the servers run $\prod_{\textnormal{MACCheck}}$ to detect malicious behaviors (line~16 in Algorithm~\ref{alg:Scheme}). If the consistency check fails, the servers immediately abort and output $\perp$, thereby ensuring the correctness and integrity of the authentication result.
The service provider reconstructs  $\eta =\Delta_{\eta} -
\langle \lambda_{\eta} \rangle_0-\langle \lambda_{\eta}  \rangle_1$  and grants the client access to the requested service if $\eta=0$; otherwise, the client's access request is rejected.

\section{Theoretical Analysis}
\label{sec:theoreticalAnalyses}

\textbf{Security analysis.} $\sysname$'s pipeline integrates a variety of maliciously-secure protocols for different computations, and the input and output of each computation in the secret-sharing domain. Using the sequential composition theorem \cite{goldreich2019play}, we deduce the overall security of $\sysname$ as stated in Theorem~\ref{theorem:sysname}. 
Due to space constraints, we present a sketch of the security proof of the $\sysname$ scheme.

\begin{theorem}
\label{theorem:sysname}
$\sysname$'s biometric authentication scheme $\prod^{\sysname}$ correctly and securely realizes the ideal functionality $\mathcal{F}^{\sysname}$ in the presence of one malicious adversary $\mathcal{A}$ in the \textnormal{(}$\prod_{\textnormal{add}}$, $\prod_{\textnormal{mult}}$, $\prod_{\mathtt{SecIP}}$, $\prod_{\mathtt{SecCMP}}$\textnormal{)}-hybrid model.
\end{theorem}

\begin{proof}[Proof Sketch]
    The malicious security of $\sysname$ is proven according to the cryptographic standard outlined in \textbf{Definition}~\ref{security}. 
    We begin by proving the security of supporting protocols ($\prod_{\textnormal{add}}$, $\prod_{\textnormal{mult}}$, $\prod_{\mathtt{SecIP}}$, $\prod_{\mathtt{SecCMP}}$) against malicious adversaries.  
    Concretely, based on SS-variants, the view of the adversary $\mathcal{A}$ in the real-world execution, denoted as $\mathsf{View}^{\prod}_{\mathcal{A}}$, is computationally indistinguishable from the simulated view $\mathsf{Sim}^{\mathcal{F}}$ generated by the ideal-world simulator $\mathsf{Sim}$.  
    To further strengthen malicious security, we integrate authentication mechanisms into these supporting protocols. Rather than relying on computationally expensive asymmetric commitment schemes, we employ a symmetric-key-based MAC to ensure integrity. The MAC authenticates the results for each secure computation, allowing any unauthorized deviations from the secure computation to be detected. If any deviations from the expected execution are detected, $\prod^{\sysname}$ aborts immediately to prevent adversarial influence. 
    Therefore, each supporting protocol is secure and correct against a PPT malicious adversary $\mathcal{A}$. 
    Finally, according to the \textit{universal composability} theory \cite{canetti2001universally}, we claim that our protocol $\prod^{\sysname}$ correctly and securely realizes the ideal functionality $\mathcal{F}^{\sysname}$ in the malicious environments. This completes the proof of Theorem~\ref{theorem:sysname}.
\end{proof} 
\begin{table}[!t]
\caption{Online theoretical performance comparison of $\sysname$ and Nomadic \cite{cheng2024nomadic}. For a fair comparison of the overall schemes, we align $\sysname$ with the over-the-threshold Nomadic scheme for one-time PPBA.  
}
\centering
\begin{adjustbox}{width=0.48\textwidth,center}
\begin{tabular}{c|c|cc} \hline\hline
&    & Computation &Communication \\ \hline 
\multirow{3}{*}{$\mathtt{SecIP}$}  & Semi-honest   &   $ {n\prod_\textnormal{mult}}$    &$nl$    \\
 & Nomadic   &   $ {2n\prod_\textnormal{mult}}$    &$2n(l+s)$  \\
  &  Ours     &  ${2\prod_\textnormal{mult}}$  &   $2(l+s)$  \\ \hline 
\multirow{3}{*}{$\mathtt{SecCMP}$}  & Semi-honest  &    ${\mathtt{Eval}^{<}}$    &  $l$ \\
& Nomadic   &   ${4\mathtt{Eval}^{<}}$    &  ${4(l+s)}$ \\
 &  Ours     &    ${\mathtt{Eval}^{<}}$    &  $2({l+s})$   \\ \hline 
\multirow{3}{*}{$\sysname$} & Semi-honest   &  ${n\prod_\textnormal{mult}+\mathtt{Eval}^{<}}$  &$(n+1)l$  \\
& Nomadic   &  ${2n\prod_\textnormal{mult}+ 4\mathtt{Eval}^{<}}$  & ${(2n+4)(l+s)}$     \\
 &  Ours     &    ${4\prod_\textnormal{mult}+\mathtt{Eval}^{<}}$  &  ${6(l+s)}$ \\  \hline \hline
\end{tabular}
\end{adjustbox} 
\vspace{-2mm}
\label{tab:comm}
\end{table}

\textbf{Efficiency analysis.} We analyze the online efficiency of $\sysname$ and its key protocols ($\mathtt{SecIP}$ and $\mathtt{SecCMP}$) against state-of-the-art (SotA) schemes in Table~\ref{tab:comm}.
%
%
There are three parameters involved in this analysis: (1) $l,s$ is the concrete parameters for $\textnormal{SPD}\mathbb{Z}_{2^l}$, (2) $n$ is the biometric template length.
Note that addition and subtraction operations are considered free in this context, as it requires no interaction and can be performed locally.
%
$\mathtt{SecIP}$, $\mathtt{SecCMP}$, and $\sysname$ show powerful improvements in online computation, with the majority of costs incurred during the offline phase for generating cryptographic primitives and randomness, such as multiplication triples and keys. Additionally, they achieve lower online communication overhead compared to SotA. 

\begin{table*}[!t]
\caption{Offline time (in minutes) and communication (Comm., in giga-bytes) of $\mathtt{SecIP}$ evaluated on vectors of length {$2^{10}$}.\label{tab:secip_offline}}
\centering
\begin{adjustbox}{width=1\textwidth,center}
\begin{tabular}{c|ccc|ccc|ccc|ccc}
\hline \hline
\multirow{2}{*}{Size} & \multicolumn{3}{c|}{LAN Time} & \multicolumn{3}{c|}{MAN Time} & \multicolumn{3}{c|}{WAN Time} & \multicolumn{3}{c}{Communication} \\
& Nomadix & Ours & Factor & Nomadix & Ours & Factor & Nomadix & Ours & Factor & Nomadix & Ours & Factor \\
\hline
$2^8$    & 0.66 & 0.64 & 1.03$\times$ & 7.64 & 7.56 & 1.01$\times$ & 12.99 & 13.07 & 0.99$\times$ & 2.34 & 2.34 & 1.00$\times$ \\
$2^{10}$ & 2.75 & 2.79 & 0.99$\times$ & 26.55 & 26.45 & 1.00$\times$ & 54.72 & 53.33 & 1.03$\times$ & 9.27 & 9.27 & 1.00$\times$ \\
$2^{12}$ & 10.77 & 10.80 & 1.00$\times$ & 106.35 & 103.56 & 1.03$\times$ & 201.41 & 200.16 & 1.01$\times$ & 37.02 & 37.02 & 1.00$\times$ \\
$2^{14}$ & 41.68 & 41.29 & 1.01$\times$ & 437.80 & 446.39 & 0.98$\times$ & 804.37 & 799.81 & 1.01$\times$ & 148.00 & 148.00 & 1.00$\times$ \\
\hline \hline
\end{tabular}
\end{adjustbox}
\end{table*}
\begin{table*}[!t]
\caption{Online time (in seconds) and communication (Comm., in mega-bytes) of $\mathtt{SecIP}$ evaluated on vectors of length {$2^{10}$}.\label{tab:secip_online}}
\centering
\begin{adjustbox}{width=1\textwidth,center}
\begin{tabular}{c|ccc|ccc|ccc|ccc}
\hline \hline
\multirow{2}{*}{Size} & \multicolumn{3}{c|}{LAN Time} & \multicolumn{3}{c|}{MAN Time} & \multicolumn{3}{c|}{WAN Time} & \multicolumn{3}{c}{Communication} \\
& Nomadix   & Ours   & Factor  & Nomadix   & Ours   & Factor  & Nomadix   & Ours   & Factor  & Nomadix     & Ours    & Factor    \\ \hline
$2^8$        & 0.39 & 0.11 & 3.64$\times$ & 0.96 & 0.23 & 4.17$\times$ & 2.22 & 0.58 & 3.83$\times$ & 4.20 & 0.004 & 1050.00$\times$ \\
$2^{10}$     & 1.47 & 0.43 & 3.45$\times$ & 3.25 & 0.68 & 4.78$\times$ & 4.95 & 1.04 & 4.76$\times$ & 16.79 & 0.016 & 1049.37$\times$ \\
$2^{12}$     & 6.09 & 1.73 & 3.52$\times$ & 15.17 & 2.48 & 6.12$\times$ & 19.25 & 3.97 & 4.85$\times$ & 67.12 & 0.063 & 1065.39$\times$ \\
$2^{14}$     & 23.60 & 7.04 & 3.35$\times$ & 65.93 & 15.65 & 4.21$\times$ & 74.73 & 20.23 & 3.69$\times$ & 268.45 & 0.25 & 1073.80$\times$ \\
\hline \hline
\end{tabular}
\end{adjustbox}
\end{table*}

\section{Experimental Evaluation}
\textbf{Testbed and parameters.} {We implement a prototype of $\sysname$ in C++. 
All experiments are conducted on two simulated servers, each equipped with an Apple M2 processor, 16 GB of RAM, running macOS Sonoma 14.4 }.
{We use the $\mathtt{tc}$ command to simulate LAN (RTT: 0.1 ms, 1 Gbps), MAN (RTT: 6 ms, 100 Mbps), and WAN (RTT: 80 ms, 40 Mbps) networks. } 
For fairness, we implement the evaluations for SotA scheme Nomadic~\cite{cheng2024nomadic} and our $\sysname$ in the MP-SPDZ framework\footnote{\url{https://github.com/data61/mp-spdz}} \cite{keller2020mp}. Since Nomadic does not release its source code, we re-implement it based on the techniques of $\textnormal{SPD}\mathbb{Z}_{{2^l}^{+}}$\cite{damgaard2019new}.
Also, the FuncSS-based comparison functions are implemented based on the library\footnote{\url{https://github.com/frankw2/libfss}}, which we enhance by incorporating the generation and evaluation algorithms from~\cite{boyle2021function}.
We set the parameters {$l=s=64$} for the $\textnormal{SPD}\mathbb{Z}_{2^l}$ secret-sharing scheme.

\textbf{Datasets and models.} We evaluate $\sysname$ on two real-world facial datasets: LFW\footnote{\url{https://www.kaggle.com/datasets/jessicali9530/lfw-dataset}} and VidTIMT \footnote{\url{https://conradsanderson.id.au/VidTIMIT/}}. 
Each biometric template is represented as a 512-dimensional feature vector ($n = 512$), extracted via FaceNet and ArcFace models, because 512 dimensions strike a practical balance between recognition accuracy and computational efficiency, and are the standard output size for these widely adopted models.
Our analysis identified 62 individuals from LFW and 43 from VidTIMT who meet the registration criteria, each with 20 valid samples.
All results are averaged over 10 trials, and the reported communication is amortized per party.

\subsection{Micro-benchmarks on Supporting Protocols}
We benchmark the offline and online costs of the building blocks for Nomadix and $\sysname$, including:
\begin{itemize}[leftmargin=3pt,topsep=0pt]
    \item[-] \textbf{Benchmark 1.} Secure inner product of vector length $2^{10}$ with varying input size.
    \item[-] \textbf{Benchmark 2.} Secure inner product of input size $2^{10}$ with varying vector lengths.
    \item[-] \textbf{Benchmark 3.} Secure comparison with varying batch size.
\end{itemize}

\begin{figure*}[!t]
    \centering
    \begin{minipage}[t]{0.24\linewidth}
        \centering
    \includegraphics[width=\textwidth]{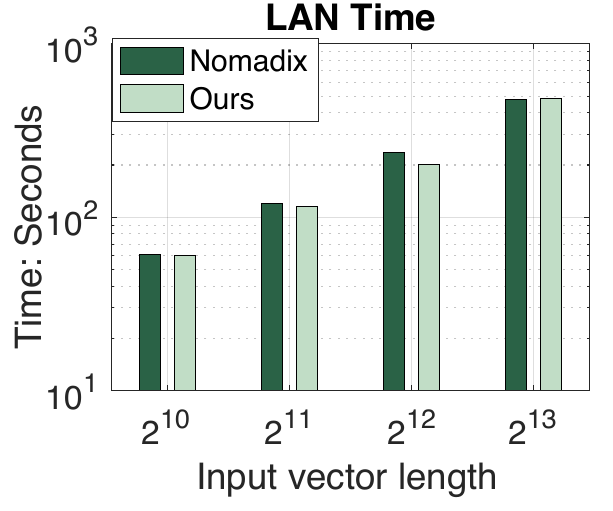}
        \vspace{-1mm} 
    \end{minipage}%
    \hfill
    \begin{minipage}[t]{0.24\linewidth}
        \centering
    \includegraphics[width=\textwidth]{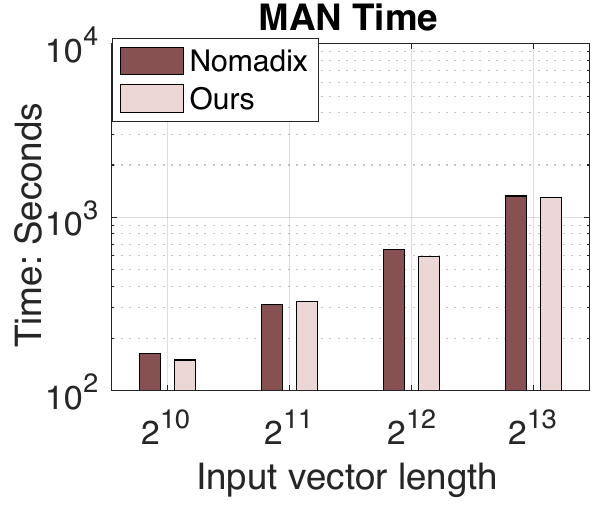} 
        \vspace{-1mm}
    \end{minipage}
     \hfill
    \begin{minipage}[t]{0.24\linewidth}
        \centering
    \includegraphics[width=\textwidth]{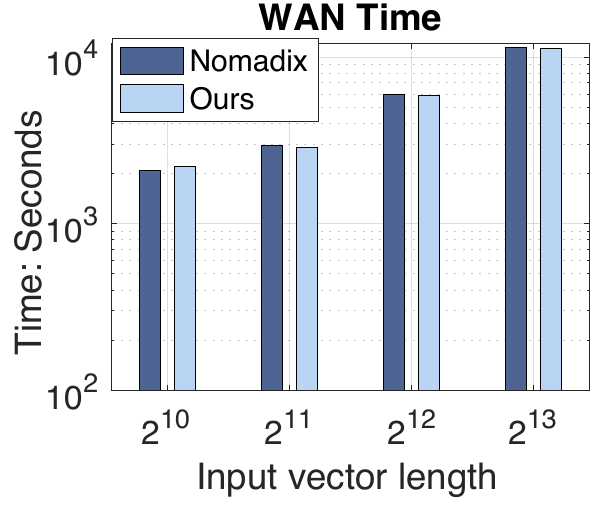} 
        \vspace{-1mm}
    \end{minipage}
     \hfill
    \begin{minipage}[t]{0.24\linewidth}
        \centering
    \includegraphics[width=\textwidth]{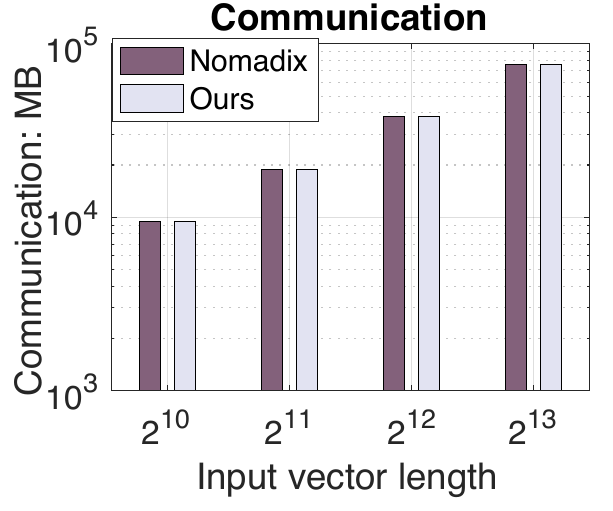} 
        \vspace{-1mm}
    \end{minipage}
    \vspace{-0.8\baselineskip}
     \caption{Offline performance comparison of $\mathtt{SecIP}$ protocol when the size of the dataset is $2^{10}$.} 
        \label{fig:secip_offline}
        \vspace{-2mm}
\end{figure*}
\begin{figure*}[!t]
    \centering
    \begin{minipage}[t]{0.24\linewidth}
        \centering
    \includegraphics[width=\textwidth]{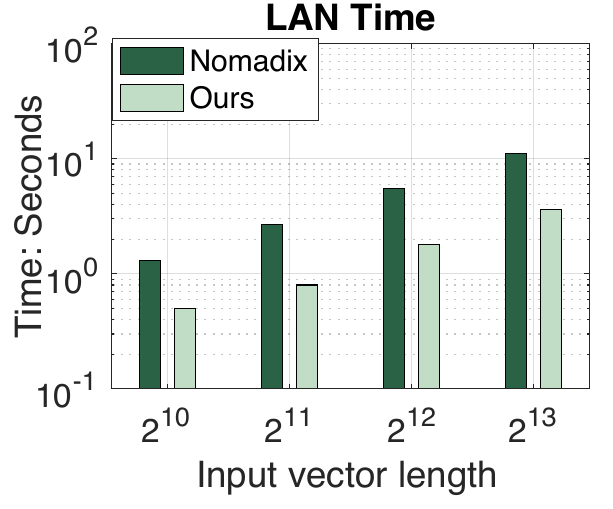}
        \vspace{-1mm} 
    \end{minipage}%
    \hfill
    \begin{minipage}[t]{0.24\linewidth}
        \centering
    \includegraphics[width=\textwidth]{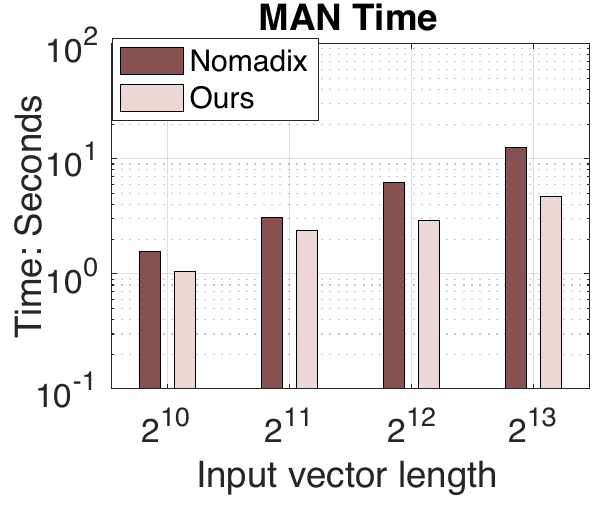} 
        \vspace{-1mm}
    \end{minipage}
     \hfill
    \begin{minipage}[t]{0.24\linewidth}
        \centering
    \includegraphics[width=\textwidth]{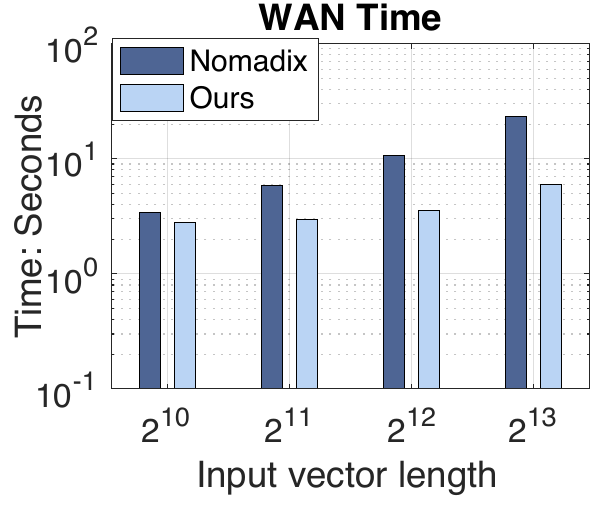} 
        \vspace{-1mm}
    \end{minipage}
     \hfill
    \begin{minipage}[t]{0.24\linewidth}
        \centering
    \includegraphics[width=\textwidth]{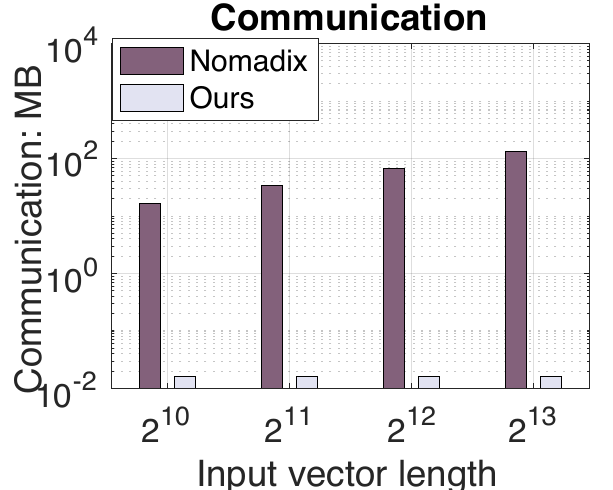} 
        \vspace{-1mm}
    \end{minipage}
    \vspace{-0.8\baselineskip}
     \caption{Online performance comparison of $\mathtt{SecIP}$ protocol when the size of the dataset is $2^{10}$.} 
        \label{fig:secip_online}
        \vspace{-2mm}
\end{figure*}

\begin{table*}[!t]
\caption{Offline time (in seconds) and communication (Comm., in mega-bytes) of $\mathtt{SecCMP}$.\label{tab:seccmp_offline}}
\centering
\begin{adjustbox}{width=1\textwidth,center}
\begin{tabular}{c|ccc|ccc|ccc|ccc}
\hline \hline
\multirow{2}{*}{Size} & \multicolumn{3}{c|}{LAN Time} & \multicolumn{3}{c|}{MAN Time} & \multicolumn{3}{c|}{WAN Time} & \multicolumn{3}{c}{Communication} \\
& Nomadix   & Ours   & Factor  & Nomadix   & Ours   & Factor  & Nomadix   & Ours   & Factor  & Nomadix     & Ours    & Factor    \\ \hline
$2^8$        &   1.43       &  0.03      &   47.66$\times$      &    3.56    &   0.12     & 29.66$\times$        &    107.31  &   4.36     &    24.61$\times$    &    221.88     &   0.69      & 321.56$\times$      \\
$2^{10}$      &  5.54       &  0.06      &   92.33$\times$     &  17.28      &  0.33      &  52.36$\times$        &    339.56   &  6.68      &   50.83$\times$      &     861.12   &   2.63      &       327.42$\times$    \\
$2^{12}$      &    21.60       &   0.19     &     113.68$\times$    &  128.89         &    1.48    &   87.08$\times$       &  577.13       &       8.24  &    70.04$\times$     &  3424.10       &      7.44   &  460.22$\times$     \\
$2^{14}$      &     86.23      &  0.65      &   132.66$\times$      &  278.53        &   3.12     & 89.27$\times$         &    2068.50      &   25.47     &  81.21$\times$       &  13668.95     &    38.41    &      355.86$\times$   \\
\hline \hline
\end{tabular}
\end{adjustbox}
\end{table*}
\begin{table*}[!t]
\caption{Online time (in seconds) and communication (Comm., in mega-bytes) of $\mathtt{SecCMP}$.\label{tab:seccmp_online}}
\centering
\begin{adjustbox}{width=1\textwidth,center}
\begin{tabular}{c|ccc|ccc|ccc|ccc}
\hline \hline
\multirow{2}{*}{Size} & \multicolumn{3}{c|}{LAN Time} & \multicolumn{3}{c|}{MAN Time} & \multicolumn{3}{c|}{WAN Time} & \multicolumn{3}{c}{Communication (MB)} \\
& Nomadix   & Ours   & Factor  & Nomadix   & Ours   & Factor  & Nomadix   & Ours   & Factor  & Nomadix     & Ours    & Factor    \\ \hline
$2^8$        & 0.16 & 0.023 & 6.96$\times$ & 0.36 & 0.05 & 7.20$\times$ & 4.30 & 0.11 & 39.09$\times$ & 0.52 & 0.02 & 28.88$\times$ \\
$2^{10}$     & 0.74 & 0.053 & 13.96$\times$ & 1.45 & 0.11 & 13.37$\times$ & 12.22 & 0.26 & 47.00$\times$ & 2.09 & 0.07 & 29.85$\times$ \\
$2^{12}$     & 2.93 & 0.18 & 16.28$\times$ & 5.77 & 0.32 & 18.03$\times$ & 15.16 & 0.54 & 28.07$\times$ & 8.71 & 0.28 & 31.10$\times$ \\
$2^{14}$     & 7.44 & 0.22 & 33.81$\times$ & 13.86 & 0.34 & 40.76$\times$ & 37.51 & 0.67 & 55.98$\times$ & 33.84 & 1.13 & 29.94$\times$ \\
\hline \hline
\end{tabular}
\end{adjustbox}
\end{table*}

\begin{table*}[!t]   
\caption{Online time (in seconds) and communication (Comm., in mega-bytes) of $\sysname$ on different biometric datasets. \label{tab:online_biometric}  
}
\centering
\begin{adjustbox}{width=1\textwidth,center}
\begin{tabular}{c|ccc|ccc|ccc|ccc}
\hline \hline
\multirow{2}{*}{Dataset} & \multicolumn{3}{c|}{LAN Time} & \multicolumn{3}{c|}{MAN Time} & \multicolumn{3}{c|}{WAN Time} & \multicolumn{3}{c}{Communication} \\
 & Nomadix   & Ours   & Factor  & Nomadix   & Ours   & Factor  & Nomadix   & Ours   & Factor  & Nomadix     & Ours    & Factor    \\ \hline
LFW &   1.78  & 0.63  &  2.82$\times$   &  3.71  &    0.85  & 4.36$\times$ &  12.78  &  1.94  &  6.58$\times$ &  12.69   &  0.13  &   97.61$\times$   \\ 
VidTIMIT &  1.28  & 0.47 &2.72$\times$   &  2.83  & 0.68 & 4.16$\times$  &  9.71  & 1.14 &  8.51$\times$  &   8.81   &  0.08  &    110.13$\times$   \\   
\hline \hline
\end{tabular}
\end{adjustbox}
\end{table*}

\textbf{Benchmark 1.} Tables~\ref{tab:secip_offline} and~\ref{tab:secip_online} report the runtime and communication costs of our $\mathtt{SecIP}$ and Nomadix under varying input sizes, for the offline and online phases, respectively.
In the offline phase (Table~\ref{tab:secip_offline}), we observe the runtime and communication costs remain nearly identical, with factor differences mostly around $1.00\times$ across all network settings.
In the online phase (Table~\ref{tab:secip_online}), $\mathtt{SecIP}$ consistently outperforms Nomadix across all metrics. We achieve substantial speedups in online runtime, with improvements ranging from $3.35\times$ to $6.12\times$ across different settings.
The communication cost in $\mathtt{SecIP}$ is reduced by more than $1000\times$ (approximately independent of the vector length $2^{10}$) compared to Nomadix, which aligns with our theoretical analysis that the online communication is independent of the vector length.

\textbf{Benchmark 2.} Figs.~\ref{tab:secip_offline} and~\ref{tab:secip_online} report the runtime and communication costs of our $\mathtt{SecIP}$ and Nomadix under varying vector lengths, for the offline and online phases, respectively. 
The results are consistent with \textbf{Benchmark 1}: Nomadix performs similarly to $\mathtt{SecIP}$ in the offline phase, with runtime and communication costs being identical for both protocols.
In the online phase, as vector size increases, Nomadix exhibits a corresponding increase in communication overhead, while $\mathtt{SecIP}$'s online communication cost remains constant. This further supports the finding that the communication cost in $\mathtt{SecIP}$ is independent of the vector length, saving about $2^{10}\times$ to $2^{13}\times$ in communication cost compared to Nomadix.

\textbf{Benchmark 3.} 
Tables~\ref{tab:seccmp_offline} (for the offline phase) and~\ref{tab:seccmp_online} (for the online phase) present the performance of the non-linear comparison protocol $\mathtt{SecCMP}$. 
In both the offline and online phases, $\mathtt{SecCMP}$ significantly outperforms Nomadix in terms of runtime and communication costs. In the offline phase, runtime improves by up to $132.66 \times$ and communication costs are reduced by up to $460.22 \times$. In the online phase, runtime improvements reach $33.81 \times$ in the LAN setting and communication costs are reduced by up to $31.10 \times$. These improvements remain consistent across different vector sizes, demonstrating $\mathtt{SecCMP}$'s efficiency.

\begin{table*}[!t]
\caption{Accuracy of $\sysname$ on different datasets and models with various choices of $k=1/ 5/ 10$.\label{tab:accuracy}}
\centering
\begin{tabular}{l|l|c|c}
\hline \hline
Datasets                  & Metrics & FaceNet & ArcFace \\ \hline
\multirow{2}{*}{\begin{tabular}[c]{@{}c@{}}{LFW} \\  {($m=1240$)}\end{tabular}} & Cosine  & 99.27\%/ 99.18\%/ 98.99\%  & 99.19\%/ 98.66\%/ 98.17\%   \\
                          & Euclidean& 99.27\%/ 99.11\%/ 98.93\%
  & 98.87\%/ 97.27\%/ 96.36\%
   \\ \hline
\multirow{2}{*}{\begin{tabular}[c]{@{}c@{}}{VidTIMIT} \\  {($m=860$)}\end{tabular}} & Cosine  & 97.91\%/ 96.14\%/ 95.76\%
    & 97.09\%/ 89.91\%/ 85.53\%
    \\
                          & Euclidean& 97.91\%/ 95.98\%/ 95.55\%
   &  96.98\%/ 88.05\%/ 81.12\%
  \\
                          \hline \hline
\end{tabular}
\end{table*}

\subsection{Evaluations on $\sysname$ Scheme}
The results presented in Table~\ref{tab:online_biometric} showcase the online performance of $\sysname$ on two different facial biometric datasets: LFW and VidTIMIT. 
For LFW, $\sysname$ achieves a remarkable factor of {2.82$\times$} faster runtime in LAN, {4.36$\times$} faster in MAN, and {6.58$\times$} faster in WAN. The communication cost is drastically reduced by a factor of {97.61$\times$}, highlighting $\sysname$'s efficiency.
Similarly, for VidTIMIT, $\sysname$ shows substantial improvements, with runtime speedup factors of 2.72$\times \sim$ 8.51$\times$ in LAN, MAN, and WAN. The communication cost is reduced by {110.13$\times$}.
These enhancements stem from our online-offline-paradigm optimized protocols with a one-round-communication design, providing strong evidence of $\sysname$'s potential for real-world biometric authentication applications.

\subsection{Accuracy}
The accuracy results of $\sysname$ with different similarity metrics are summarized in Table~\ref{tab:accuracy}. 
Top-$k$ accuracy is calculated as the number of samples whose true label appears in the top-$k$ predicted labels, divided by the total number of samples evaluated.
 We can see that FaceNet, which is used for biometric feature extraction, consistently outperforms ArcFace in all cases ($k = 1, 5, 10$), achieving accuracy rates exceeding $95\%$. Considering both biometric authentication accuracy, we use FaceNet as the source of all real biometric features in the above experiments. Also, Cosine similarity performs better than Euclidean distance in facial authentication, providing valuable insights for future research on selecting appropriate biometric similarity metrics.

\section{Conclusion and Future Work}

In this paper, we present $\sysname$, a 2PC-based biometric authentication scheme that supports flexible similarity metrics, including cosine similarity and Euclidean distance.
Through tailored protocols with an offline-online paradigm, $\sysname$ ensures low-latency authentication while providing strong privacy and integrity guarantees against malicious adversaries. Finally, experimental results on real-world biometric datasets demonstrate the practicality and effectiveness of $\sysname$, highlighting its suitability for real-time, large-scale biometric authentication scenarios.

In future work, we aim to extend $\sysname$ to support multi-modal biometric authentication, integrating multiple biometric modalities such as fingerprint, face, and audio recognition to further enhance accuracy, robustness, and security. 
We also plan to explore the design of an end-to-end PPBA schemes, including privacy-preserving de-duplication during enrollment, to further improve practicality and address real-world deployment challenges.

\section*{Acknowledgments}
This research is supported by the National Research Foundation, Singapore and Infocomm Media Development
Authority under its Trust Tech Funding Initiative (DTC-T2FI-CFP-0002). Any opinions, findings and conclusions or recommendations expressed in this material are those of the author(s) and do not reflect the views of National Research Foundation, Singapore and Infocomm Media Development Authority.

\bibliographystyle{splncs04}
\bibliography{mybib}

\appendix

\textbf{Security Proof.}
In this section, we first provide simulation-based security proofs for the supporting protocols. Then, we present the ideal-world definition of our proposed $\sysname$ (see Figure~\ref{fig:IdealFunctionality-Auth}) and give a hybrid argument proof for the security of the $\sysname$ scheme.

\begin{theorem}
\label{theo3}
In the $(\mathcal{F}_{\textnormal{ABB}}, \prod_{\textnormal{add}}, \prod_{\textnormal{mult}})$-hybrid model, the protocol $\prod_{\mathtt{SecIP}}$ implements $\mathcal{F}_{\mathtt{SecIP}}$ correctly and securely against malicious adversary.
\end{theorem}
\begin{proof} 
The proof of the correctness is established in Theorem~\ref{theo1}. We construct an ideal-world simulator $\mathsf{Sim}_{\mathtt{SecIP}}$ as follows to show that the real $\prod_{\mathtt{SecIP}}$ securely realizes the functionality $\mathcal{F}_{\mathtt{SecIP}}$ in the presence of a malicious adversary $\mathcal{A}$.

\begin{enumerate}[leftmargin=*,topsep=0pt]
   \item $\mathsf{Sim}_{\mathtt{SecIP}}$ receives the public parameters $l,s$ and the shares $\langle \lambda_X \rangle,\langle \lambda_{\phi X} \rangle,\langle \lambda_Y \rangle \leftarrow \mathbb{Z}_{l+s}^n$ from the ideal functionality $\mathcal{F}_{\mathtt{SecIP}}$.
    
    \item $\mathsf{Sim}_{\mathtt{SecIP}}$ simulates the ideal functionalities $\mathcal{F}_{\textnormal{rand}}$ and $\mathcal{F}_{\textnormal{triple}}$. It randomly samples:
    \[
     A_1', A_2', B', C_1', C_2' \leftarrow \mathbb{Z}_{l+s}^n,  \quad \lambda_{z}' ,\lambda_{\phi z}'\leftarrow \mathbb{Z}_{l+s}^1,\]
    and generates corresponding additive shares $ \langle A_1'\rangle, \langle A_2'\rangle, \langle B'\rangle, \langle C_1'\rangle, \langle C_2'\rangle,\langle \lambda_{z}'\rangle,\langle \lambda_{\phi z}'\rangle$, and sends these simulated shares to the malicious adversary. 

    \item When the malicious adversary computes the differences: $\langle \delta_X' \rangle =\langle A_1' \rangle - \langle \lambda_X' \rangle,
    \langle \delta_{\phi X}' \rangle = \langle A_2'\rangle - \langle \lambda_{\phi X}'\rangle,
    \langle \delta_Y' \rangle= \langle B'\rangle - \langle \lambda_Y'\rangle$, and sends the shares of $\langle \delta_X' \rangle$, $\langle \delta_{\phi X}' \rangle$, and $\langle \delta_Y' \rangle$ to open. $\mathsf{Sim}_{\mathtt{SecIP}}$ sends the clear values of $\delta_X'$, $\delta_{\phi X}'$, and $\delta_Y'$ to the adversary.
    
    \item Upon receiving the adversary’s optimized shares: $(\Delta_{X}, \langle \lambda_X\rangle ),(\Delta_{\phi X}, \langle \lambda_{\phi X}\rangle ),(\Delta_{Y},\langle \lambda_Y\rangle ),\mathsf{Sim}_{\mathtt{SecIP}}$ computes: $\Delta_{z}' = \sum_{i=1}^{n}( (\Delta_X[i]+\delta_X'[i])\left(\Delta_Y[i]+\delta_Y'[i]\right)- A_1'[i] (\Delta_Y[i]+\delta_Y'[i])-\left(\Delta_X[i]+\delta_X'[i]\right) B'[i]  + C_1'[i] ) + \lambda_{z}'$ and $\Delta_{\phi z}' = \sum_{i=1}^{n}( (\Delta_{\phi X}[i]+\delta_{\phi X}'[i])\left(\Delta_Y[i]+\delta_Y'[i]\right)- A_2'[i] \left(\Delta_Y[i]+\delta_Y'[i]\right)-(\Delta_{\phi X}'[i]+\delta_{\phi X'}[i]) B'[i]  + C_2'[i] ) + \lambda_{\phi z}'$.
    
    \item $\mathsf{Sim}_{\mathtt{SecIP}}$ opens $\Delta_{z}'$ and$\Delta_{\phi z}'$ and returns them to the adversary. 

    \item $S_{\mathtt{SecIP}}$ outputs all values the adversary would have seen in the real protocol, $\mathsf{Sim}_{\mathtt{SecIP}}$ verifies the consistency of the shares via MAC checks. If the MAC checks fail, $\mathsf{Sim}_{\mathtt{SecIP}}$ aborts and notifies $\mathcal{F}_{\mathtt{SecIP}}$ to abort as well.
\end{enumerate} 

Since the simulator $\mathsf{Sim}_{\mathtt{SecIP}}$ only uses simulated values and the distributions of these shares are indistinguishable from those in the real protocol due to the information-theoretic security of secret sharing primitives, the view $\mathsf{View}_{\mathtt{SecIP}}$ of the adversary $\mathcal{A}$ in the simulation is indistinguishable from its view in the real execution. Moreover, any deviation from the protocol $\prod_{\mathtt{SecIP}}$ is detected through $\prod_{\textnormal{MACCheck}}$ and leads to an abort, just as in the real world. Therefore, the real protocol $\prod_{\mathtt{SecIP}}$ securely realizes the ideal functionality $\mathcal{F}_{\mathtt{SecIP}}$.
\end{proof}

\begin{theorem}
\label{theo4}
In the $(\mathcal{F}_{\textnormal{ABB}}, \prod_{\mathtt{Gen}},\prod_{\mathtt{Eval}})$-hybrid model, the protocol $\prod_{\mathtt{SecCMP}}$ implements $\mathcal{F}_{\mathtt{SecCMP}}$ correctly and securely against malicious adversary.
\end{theorem}
\begin{proof}
The proof of the correctness is established in Theorem~\ref{theo2}. We construct an ideal-world simulator $S_{\mathtt{SecCMP}}$ to simulate the view $\mathsf{View}_{\mathtt{SecCMP}}$ of a malicious adversary $\mathcal{A}$ in $\prod_{\mathtt{SecCMP}}$ and show that it securely realizes the ideal functionality $\mathcal{F}_{\mathtt{SecCMP}}$.

\begin{enumerate}[leftmargin=*,topsep=0pt]
    \item $\mathsf{Sim}_{\mathtt{SecCMP}}$ receives the the public parameters $l,s$, input shares $\langle \lambda_x \rangle$, $\langle \phi \rangle \leftarrow \mathbb{Z}_{l+s}$ from the ideal functionality $\mathcal{F}_{\mathtt{SecCMP}}$.

    \item $\mathsf{Sim}_{\mathtt{SecCMP}}$ opens $a = \lambda_x$ and $\textbf{b} = (1, \phi)$.

    \item $\mathsf{Sim}_{\mathtt{SecCMP}}$ runs the (honest) $\mathtt{Gen}_{a, -\textbf{b}}^{<}$ algorithm to get $(\kappa_0'', \kappa_1'')$, sets $\kappa_{\theta}''' = \kappa_{\theta}'' \| \langle \textbf{b} \rangle_{\theta}$, then sends $\kappa_{\theta}'''$ to the adversary.  

    \item $\mathsf{Sim}_{\mathtt{SecCMP}}$ receives the invocations to $\mathcal{F}_\textnormal{share}$ from the adversary $\mathcal{A}$, returns shares $\langle b_0' \rangle,\langle b_1' \rangle$ to $\mathcal{A}$.  

    \item $\mathsf{Sim}_{\mathtt{SecCMP}}$ receives the $\mathcal{F}_{\textnormal{rand}}$ calls from $\mathcal{A}$, randomly generates corresponding additive shares $ \langle \lambda_{z}'\rangle,\langle \lambda_{\phi z}'\rangle$, then sends these simulated shares to the adversary. 

    \item Upon receiving the adversary's optimized share $(\Delta_x, \langle \lambda_x \rangle)$, $\mathsf{Sim}_{\mathtt{SecCMP}}$ runs the honest evaluation algorithm: $(\langle \gamma_0' \rangle, \langle \gamma_1' \rangle) \leftarrow \mathtt{Eval}_{a, -\textbf{b}}^{<}(\theta, \kappa_{\theta}', \Delta_x)$.

    \item $\mathsf{Sim}_{\mathtt{SecCMP}}$ learns $\langle \Delta_z' \rangle = \langle \gamma_0' \rangle + \langle b_0' \rangle + \langle \lambda_z' \rangle$, $\langle \Delta_{\phi z}' \rangle = \langle \gamma_1' \rangle + \langle b_1' \rangle + \langle \lambda_{\phi z}' \rangle$.

    \item $\mathsf{Sim}_{\mathtt{SecCMP}}$ opens $\Delta_z'$ and $\Delta_{\phi z}'$ to the adversary.

    \item $\mathsf{Sim}_{\mathtt{SecCMP}}$ simulates $\prod_{\textnormal{MACCheck}}$ using all the adversary's values. If the check fails, abort the execution with the adversary $\mathcal{A}$; otherwise, continue.
\end{enumerate}

Since all values are either correctly computed using values from the ideal functionality or uniformly random from the adversary's perspective, and any deviation is caught by the MAC check, the simulator's $\mathsf{Sim}_{\mathtt{SecCMP}}=\{\kappa_{\theta}''', \langle \lambda_z' \rangle,\langle \lambda_{\phi z}' \rangle , \langle b_0' \rangle, \langle b_1' \rangle,\langle \gamma_0' \rangle, \langle \gamma_1' \rangle,\Delta_z', \Delta_{\phi z}'\}$ is indis- tinguishable from that in the real view $\mathsf{View}_{\mathtt{SecCMP}}=\{\kappa_{\theta}$, $\langle \lambda_z \rangle,\langle \lambda_{\phi z} \rangle ,\langle b_0 \rangle, \langle b_1 \rangle, \langle \gamma_0 \rangle, \langle \gamma_1 \rangle,  \Delta_z, \Delta_{\phi z}\}$. Thus, $\prod_{\mathtt{SecCMP}}$ securely realizes $\mathcal{F}_{\mathtt{SecCMP}}$.
\end{proof}

\begin{figure}[!ht]
\centering
\begin{center} 
\fbox{
\begin{varwidth}{0.46\textwidth}
 \item[-] \textbf{Parameters}. The servers $\mathcal{P}_0$ and $\mathcal{P}_1$ has, security setings $\{l,s\}$,  pre-stored the authenticated MAC key $[\![ \phi ]\!]$ and the shared biometric database (i.e., identities and biometric templates) $  [\![ \vec{\textbf{D}} ]\!]  = [[\![I ]\!], [\![\widehat{\textbf{D}} ]\!]]$.

 \item[-] \textbf{Input}. The client $\mathcal{C}$ sends secret shares of the fresh pre-processed biometric template $[\![ \widehat{T} ]\!]$ with identity index $[\![ I_{\mathcal{C}}  ]\!]$ to $\mathcal{F}^{\sysname}$.
 
 \item[-] \textbf{Computation}. $\mathcal{F}^{\sysname}$ internally computes the similarity scores between $[\![ \widehat{T} ]\!]$ and all stored templates $[\![ \widehat{\textbf{D}}[i] ]\!]$ using $\prod_{\mathtt{SecIP}}$, producing shared similarity results $[\![S[i]]\!], [\![\phi S[i]]\!]$ for all $i \in [1, m]$. Then, it identifies the top-1 match using $\prod_{\mathtt{SecCMP}}$, and obtains the result $[\![\eta]\!], [\![\phi \eta]\!]$.

 \item[-] \textbf{Integrity Check}. $\mathcal{F}^{\sysname}$ runs $\prod_{\textnormal{MACCheck}}([\![\eta]\!], [\![\phi \eta]\!])$. If the MAC check fails, $\mathcal{F}^{\sysname}$ aborts and outputs $\perp$.

 \item[-] \textbf{Output}. If the MAC check passes, $\mathcal{F}^{\sysname}$ returns the authenticated result $[\![\eta]\!]$ to the service provider $\mathcal{S}$ and nothing to the client $\mathcal{C}$. $\mathcal{S}$ grants access if $\eta = 0$.
\end{varwidth}
}
\vspace{-1mm}
\end{center}
\caption{{Ideal Functionality $\mathcal{F}^{{\sysname}}$.} } 
\vspace{-2mm}
\label{fig:IdealFunctionality-Auth}
\end{figure}
 
Next, we present the security proof for Theorem~\ref{theorem:sysname}, where $\mathcal{P}_0$ and $\mathcal{P}_1$ access to the simulator for ideal functionality of the supporting protocols $\prod_\textnormal{OptSS}$, $\prod_\mathtt{SecIP}$, $\prod_\mathtt{SecCMP}$, $\prod_\textnormal{mult}$, and $\prod_\textnormal{MACCheck}$ of the $\sysname$ scheme in the following specified order.

\noindent $\textbf{Hyb}_0$: $\mathcal{P}_0$ and $\mathcal{P}_1$ jointly execute the real $\prod_\textnormal{OptSS}$ protocol to obtain the initial OptSS shares of the input data. All subsequent protocols, including $\prod_\mathtt{SecIP}$, $\prod_\mathtt{SecCMP}$, $\prod_\textnormal{mult}$, and $\prod_\textnormal{MACCheck}$, are replaced by their respective ideal functionalities accessed via the simulator. Since the ideal functionalities leak nothing beyond what is revealed by the output, the malicious adversary's view in this hybrid is determined solely by the execution of $\prod_\textnormal{OptSS}$, which have been proven in \cite{ben2019turbospeedz}. Therefore, the view of any PPT adversary in $\textbf{Hyb}_0$ is simulatable, and the adversary learns nothing beyond the prescribed output.

\noindent $\textbf{Hyb}_1$: Building upon $\textbf{Hyb}_0$, $\mathcal{P}_0$ and $\mathcal{P}_1$ execute both the real $\prod_\textnormal{OptSS}$ and $\prod_\mathtt{SecIP}$ protocols, while the functionalities of $\prod_\mathtt{SecCMP}$, $\prod_\textnormal{mult}$, and $\prod_\textnormal{MACCheck}$ remain ideal and are realized via the simulator. The difference between the views of $\mathcal{P}_0$ and $\mathcal{P}_1$ in $\textbf{Hyb}_1$ and $\textbf{Hyb}_0$ lies only in the substitution of the ideal functionality of $\prod_\mathtt{SecIP}$ with its real execution. Since $\prod_\mathtt{SecIP}$ is proven against the malicious adversary in Theroem~\ref{theo3}, its view is simulatable, and thus no PPT adversary can distinguish $\textbf{Hyb}_1$ from $\textbf{Hyb}_0$.

\noindent $\textbf{Hyb}_2$: Further to $\textbf{Hyb}_1$, $\mathcal{P}_0$ and $\mathcal{P}_1$ execute the real $\prod_\mathtt{SecCMP}$ protocol in addition to the real executions of $\prod_\textnormal{OptSS}$ and $\prod_\mathtt{SecIP}$, while the remaining protocols $\prod_\textnormal{mult}$ and $\prod_\textnormal{MACCheck}$ are still implemented via ideal functionalities. As the only difference between $\textbf{Hyb}_2$ and $\textbf{Hyb}_1$ is the real execution of $\prod_\mathtt{SecCMP}$ instead of its ideal functionality, and since $\prod_\mathtt{SecCMP}$ is also simulatable (Theorem~\ref{theo4}), it follows that no PPT adversary can distinguish $\textbf{Hyb}_2$ from $\textbf{Hyb}_1$.

\noindent $\textbf{Hyb}_3$: On the basis of $\textbf{Hyb}_2$, $\mathcal{P}_0$ and $\mathcal{P}_1$ additionally execute the real $\prod_\textnormal{mult}$ protocol, while only $\prod_\textnormal{MACCheck}$ is still realized by its ideal functionality via the simulator. Again, the only difference between the adversarial view in $\textbf{Hyb}_3$ and $\textbf{Hyb}_2$ is due to replacing the ideal version of $\prod_\textnormal{mult}$ with its real protocol. As shown above, $\prod_\textnormal{mult}$ is malicious secure \cite{ben2019turbospeedz,yuan2024md}, so the adversary's view remains simulatable. Hence, $\textbf{Hyb}_3$ and $\textbf{Hyb}_2$ are indistinguishable.

\noindent $\textbf{Hyb}_4$: Finally, $\mathcal{P}_0$ and $\mathcal{P}_1$ execute the full $\sysname$ scheme, including the real execution of $\prod_\textnormal{OptSS}$, $\prod_\mathtt{SecIP}$, $\prod_\mathtt{SecCMP}$, $\prod_\textnormal{mult}$, and $\prod_\textnormal{MACCheck}$ protocols. The only difference between $\textbf{Hyb}_4$ and $\textbf{Hyb}_3$ lies in the replacement of the ideal version of $\prod_\textnormal{MACCheck}$ with its real implementation. As the $\prod_\textnormal{MACCheck}$ protocol is proven to be malicious secure \cite{bai2023mostree} and its view simulatable, $\textbf{Hyb}_4$ is indistinguishable from $\textbf{Hyb}_3$.

In the above real protocol executions, any additive error introduced by a malicious adversary will cause the MAC check to fail, except with probability at most $\frac{1}{2^{l+s}-1}$. Therefore, the simulation remains statistically indistinguishable from the real execution, up to a negligible statistical error. This establishes the security of the $\sysname$ scheme against malicious adversaries.

\end{document}